\documentclass[english]{iopart}
\usepackage[T1]{fontenc}
\usepackage[latin1]{inputenc}
\usepackage{color}
\usepackage{graphicx}
\usepackage{amssymb}

\makeatletter

 \usepackage{graphicx}
 \usepackage{cite}

\usepackage{babel}
\makeatother
\begin{document}

\jl{19}

\paper[The transfer of energy between electrons and ions in solids]{The transfer of energy between electrons and ions in solids}

\author{A. P. Horsfield$^{1}$, D. R. Bowler$^{1,2}$, H. Ness$^{3}$, C.
G. S\'{a}nchez$^{4}$, T. N. Todorov$^{5}$, and A. J. Fisher$^{1,2}$}

\address{$^{1}$Department of Physics and Astronomy, University College London,
Gower Street, London WC1E 6BT, United Kingdom}

\address{$^{2}$London Centre for Nanotechnology, University College London,
Gower St, London WC1E 6BT}

\address{$^{3}$Service de Physique et de Chimie des Surfaces et Interfaces,
DSM/DRECAM, CEA-Saclay, 91191 Gif sur Yvette, France}

\address{$^{4}$Unidad de Matem\'atica y F\'{\i}sica, Facultad de Ciencias
Qu\'{\i}micas, INFIQC, Universidad Nacional de C\'ordoba, Ciudad
Universitaria, 5000 C\'ordoba, Argentina}

\address{$^{5}$School of Mathematics and Physics, Queen's University of Belfast,
Belfast BT7 1NN, United Kingdom}

\eads{a.horsfield@ucl.ac.uk, david.bowler@ucl.ac.uk, t.todorov@qub.ac.uk,
ness@dsm-mail.saclay.cea.fr}

\maketitle
\begin{abstract}
In this review we consider those processes in condensed matter that
involve the irreversible flow of energy between electrons and nuclei
that follows from a system being taken out of equilibrium. We survey
some of the more important experimental phenomena associated with
these processes, followed by a number of theoretical techniques for
studying them. The techniques considered are those that can be applied
to systems containing many non-equivalent atoms. They include both
perturbative approaches (Fermi's Golden Rule, and non-equilibrium
Green's functions) and molecular dynamics based (the Ehrenfest approximation,
surface hopping, semi-classical gaussian wavefunction methods and
correlated electron-ion dynamics). These methods are described and
characterised, with indications of their relative merits.
\end{abstract}

\section{Introduction}

This review is about the transfer of energy between electrons and
nuclei. We focus especially on the theories and computer models that
can be used to describe this process. Energy transfer is such a general
concept, however, that we need to specify what exactly we have in
mind here. A possible sequence of events that illustrates the kind
of processes we have in mind is:

\begin{enumerate}
\item A system begins in equilibrium.
\item It is then subjected to some external influence. This might be an
electromagnetic pulse that transfers energy to the electrons thousands
of times more efficiently than to the nuclei, thus giving the electrons
a relatively higher energy. Or, at the other extreme, it might be
a flux of neutrons which interact only with the nuclei which would
have the effect of raising the energy of the nuclei relative to the
electrons. (It could also be a number of other things, such as a beam
of electrons or a flux of heat.)
\item As a result of this departure from equilibrium, there will be a net
flow of energy either from the electrons to the nuclei, or from the
nuclei to the electrons, as the system moves back towards equilibrium.
\end{enumerate}
The final transfer of energy is made possible by interactions that
couple the electrons to the nuclei. Of course these same interactions
lead to other forms of correlated motion between electrons and nuclei,
such as the modification of the electronic band structure, the effective
coupling between electrons that produces superconductivity, and the
coherent transport of electrons by virtual polarons. Interesting though
they are they are not covered here.

This review is structured as follows. In the next section we briefly
summarise some relevant phenomenology. We then describe the most important
methods currently available for modelling these phenomena. Finally
we briefly consider the future of the theories.

\section{Phenomenology}

In this section we survey some phenomena observed in solids that are
produced by the irreversible flow of energy between electrons and
nuclei. \emph{}In what follows, neither the range of phenomena nor
the list of citations are exhaustive. Instead we have merely attempted
to provide a basis for thought and analysis for the interested reader.

\subsection{Joule heating}

Probably the most familiar and easiest to observe phenomenon involving
the inelastic transfer of energy between electrons and nuclei is the
increase in temperature of an electrical conductor when a current
passes through it. The idea that electric currents do work on conductors,
and so cause heating, has been known since the mid nineteenth century
when it was investigated by James Joule\cite{joule-1845-a}. The treatment
of transport in terms of Boltzmann's equation has been brought to
a high level\cite{ziman-1960-a}, with the coupling between the electrons
and phonons treated perturbatively. This is accurate for macroscopic
materials because no coupling to any one vibrational mode is large.

With the rise in interest in mesoscale and nanoscale systems additional
quantum effects became apparent, introduced by the small length scales.
One such system is the atomic scale contact, which can be produced
either by a break junction or a scanning tunnelling microscope. When
a current is passed through the junction heating takes place that
increases with applied voltage. Experimental evidence for this heating
comes from observations of the voltage dependence of two level fluctuations
and hysteresis at conductance discontinuities in atomic scale contacts\cite{muller-1992-a,vandenbrom-1998-a}
and from measurements on current induced rupture of atomic chains\cite{smit-2004-a}.
This heating, which can be very substantial, may at first seem mysterious
since the electron mean free path is so much greater than an atomic
spacing. However this net heating is simply a compromise between the
small probability for an individual electron to scatter off a phonon
in an atomic wire, and the huge density of current carrying electrons
that accompanies the current densities attainable in quasi-ballistic
metallic nanoconductors\cite{ralls-1989-a,holweg-1992-a,todorov-1998-a}.
This can result in the apparent paradox of a hot conductor that is
still largely ballistic, and thus resistance free, as far as individual
electrons coming through are concerned.

\subsection{Relaxation of excited electrons}

The transfer of energy is not the only important consequence of the
non-adiabatic interactions between electrons and nuclei. For example,
non-radiative processes in semiconductors contribute to trapping of
free carriers which has an impact on the conductivity and optical
properties\cite{stoneham-1981-a,stoneham-2001-a}. Because the binding
energy of deep centres can be much larger than typical phonon energies,
the phonons can participate in the optical transitions (which provide
the most direct information about the defect electronic properties).
Thus non-radiative processes can be observed in optical absorption
spectra of lattice defects\cite{boer-2002-a}.

Paramagnetic ions in insulators can undergo spin reversal\cite{standley-1969-a}
(the ions switch between Zeeman levels). Initially an electromagnetic
field is applied. This takes the population away from equilibrium.
The system can then return to equilibrium through interaction with
thermal phonons\cite{stevens-1967-a}, a process that can be observed
by spin resonance experiments\cite{brya-1967-a}. There is intense
interest at present in similar phenomena in nanoscale systems such
as quantum dots, as the quantum confinement produces discrete electronic
states which modify the relaxation rates of electrons and holes\cite{gundogdu-2005-a,buyanova-2005-a,lu-2005-a} 

Excitons can be thought of as a bound pair of particles: a negatively
charged electron and a positively charged hole. Excitons can further
interact with each other and form pairs\cite{haynes-1966-a}, or bind
to defects\cite{choyke-1962-a,gilleo-1968-a}. However, since the
hole is simply the absence of an electron, with the positive charge
originating with the nuclei, it is possible for the two quasi-particles
to combine and annihilate with the release of energy. This energy
can escape as a photon or one or more phonons. In semiconductors a
phonon may be essential to the recombination process because of crystal
momentum conservation, though the presence of defects that trap excitons
can remove this constraint\cite{choyke-1962-a}. The angular momentum
state of the exciton can frustrate its radiative decay\cite{gilleo-1968-a},
which is important for the efficiency of organic light-emitting diodes.
A simple counting argument suggests roughly a quarter of excitons
are in a singlet state (and can undergo radiative decay) while the
remaining three quarters are in a triplet state (and cannot undergo
radiative decay). However, there appear to be significant exceptions
to this rule\cite{wohlgenannt-2001-a}. Excitons in quantum wells
have increased binding energy because of their confinement, though
their interaction with phonons appears unaffected\cite{feldmann-1978-a}.
In multiple quantum well structures, excitons can become trapped at
a number of sites with different energies. Phonons assist in the migration
at low temperatures (making up the difference in energies between
binding sites), and at higher temperatures ($>$10K) produce thermally
activated migration\cite{wang-1990-a}.

\subsection{Inelastic electron tunnelling spectroscopy}

\begin{figure}
\begin{center}\includegraphics[%
  width=0.8\columnwidth,
  keepaspectratio]{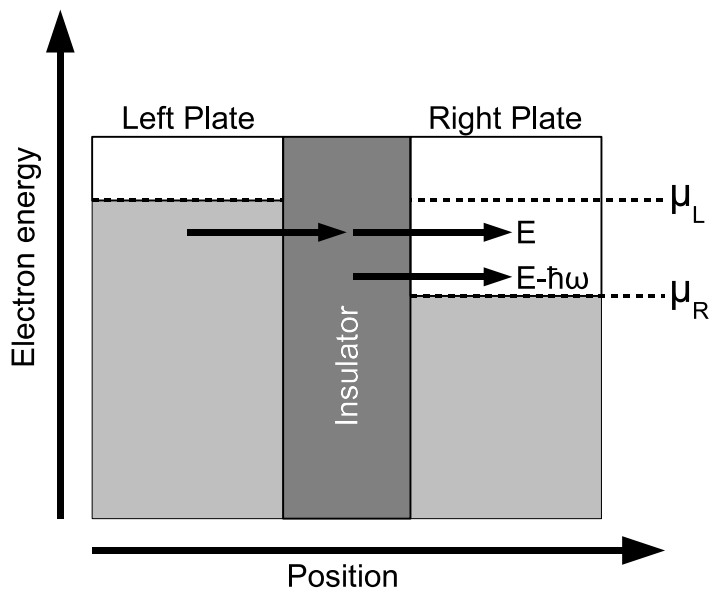}\end{center}

\caption{\label{fig:iets01} The arrangement for inelastic electron tunneling
spectroscopy. There are two metallic plates, between which is an insulating
layer. A voltage is applied, so the electron chemical potentials of
the two plates become offset, allowing electrons to flow from filled
states on the left to empty states on the right. In the insulating
sample there are molecules that can exchange energy with a tunneling
electron: the electron can excite a vibration of angular frequency
$\omega$ in the molecule, and so lose energy $\hbar\omega$.}
\end{figure}

Inelastic electron tunnelling spectroscopy exploits the transfer of
energy between electrons and nuclei to probe the vibrational spectrum
of molecules by passing an electric current through them. A schematic
of the experimental arrangement is shown in figure \ref{fig:iets01}.
There are two metallic plates (or contacts) across which a bias is
applied ($eV=\mu_{L}-\mu_{R}$, where $V$ is the applied voltage
and $\mu_{L}$ and $\mu_{R}$ are the chemical potentials of the left
and right plates respectively, as shown in figure \ref{fig:iets01}).
Between the plates is an insulating region (often an oxide) that contains
the molecules whose vibrational frequencies we wish to probe. In general
there will be an elastic tunnelling current (an electron arrives at
the right plate with the same energy that it had when it departed
the left plate) whose magnitude increases linearly with the applied
voltage for low voltages. At low temperatures, and for sufficiently
large applied voltage ($eV\ge\hbar\omega_{0}$ where $\omega_{0}$
is the angular frequency for the vibrational mode with lowest frequency
that can exchange energy with an electron), a conduction channel will
open up in which an electron can arrive at the right hand plate and
enter an empty state with a diminished energy, the energy lost being
equal to one quantum of vibrational energy for the molecules in the
sample. By increasing the voltage, additional channels open up corresponding
to higher vibrational frequencies, or possibly multi-phonon processes.
At higher temperatures it is possible for the electron to gain energy
from the molecular oscillations.

Lambe and Jaklevic\cite{lambe-1968-a} present a theory based on Fermi's
Golden Rule (see section \ref{sub:Perturbation-theory}) suitable
for oxides in which the electron exchanges energy remotely from the
oscillators through an electrostatic dipole interaction. They also
report experimental results for a system consisting of Al and Pb plates
sandwiching ${\rm Al}_{2}{\rm O}_{3}$ doped with small organic molecules.
The experiments were performed from below 1K up to 300K, and inelastic
resonances corresponding to molecular vibrational frequencies were
detected as peaks in the second derivative of the current-voltage
curve (${\rm d}^{2}I/{\rm d}V^{2}$). Klein \emph{et al}\cite{klein-1973-a}
used a similar technique, and found that different ranges of voltage
probed not only the frequencies of the impurities but also of the
contacts and the oxide. The technique for absorbing molecules was
improved by Simonsen \emph{et al}\cite{simonsen-1974-a} who used
the liquid phase of the molecules. This made it possible for them
to investigate the vibrational frequency spectrum of much larger molecules
including proteins and nucleotides. Kirtley and co-workers have made
a series of contributions\cite{kirtley-1975-a,kirtley-1976-a,kirtley-1976-b,kirtley-1979-a,kirtley-1980-a}.
They have investigated the effect of the choices of metal and oxide
for the tunnelling device, and found that these can influence the
measured spectra\cite{kirtley-1975-a,kirtley-1976-b}. They have developed
a theory of the intensities of the vibrational modes using a transfer-Hamiltonian
formalism with WKB wavefunctions\cite{kirtley-1976-a}. In this model
the interaction between the tunnelling electron and an oscillating
molecule is through the Coulomb interaction with the partial charges
on the molecule. It has been applied to ${\rm CH}_{3}{\rm SO_{3}^{-}}$
chemisorbed on alumina\cite{kirtley-1980-a}. Multiple scattering
X$\alpha$ calculations showed that this long ranged potential indeed
dominates the measured spectra\cite{kirtley-1979-a}.

Persson and Baratoff\cite{persson-1987-a} showed that the excitation
and immediate de-excitation of a molecular vibration by a tunneling
electron can give a decrease in the resonant conductance. A closely
related problem is the inelastic current voltage spectroscopy of a
ballistic atomic metallic chain\cite{agrait-2002-a,agrait-2002-b},
or a single resonant molecule between two electrodes\cite{smit-2002-a}.
The opening of new inelastic scattering channels, as the bias matches
the energies of various phonon modes in the wire, also leads to a
suppression of the electronic conductance, since, starting from ideal
transmission, the conductance can only ever go down, and the inelastic
I-V features take the form of dips in \emph{}${\rm d}^{2}I/{\rm d}V^{2}$.

Scanning tunnelling microscopes (STMs) have been used for inelastic
spectroscopy. Inspired by this, Gregory\cite{gregory-1990-a} created
a very small junction by using two crossing gold wires with argon
and carbon monoxide in between. Resonant peaks were observed from
hydrocarbon contaminants, as was Coulomb blockade behaviour. Ho and
co-workers\cite{stipe-1998-a,stipe-1998-b,stipe-1999-a,hahn-2000-a}
have used an STM to probe the orientation, motion and vibrational
spectra of small molecules on metal surfaces. Yu \emph{et al.} studied
a molecular transistor and found shifts in the resonances when the
gate voltage was varied. With the experimental advances has gone developments
in the theory. Lorente and Persson\cite{lorente-2000-a} used density
functional theory and a perturbative implementation of non-equilibrium
Green's functions (NEGF) to study ${\rm C}_{2}{\rm H}_{2}$ and ${\rm C}_{2}{\rm HD}$
on Cu(100), and Galperin \emph{et al}.\cite{galperin-2004-a} used
NEGF and a model Hamiltonian to study line shapes and line widths
of IETS features.

\subsection{Friction force on high energy particles}

The bombardment of materials by high energy heavy particles occurs
in a number of situations. Ions with energies of order 10 keV are
used for ion implantation in semiconductors\cite{williams-1998-a},
while in hydrogen fusion powerplants materials of quite extraordinary
resilience are required to surround the region containing the plasma
because they are subjected to bombardment by highly energetic neutrons
(14.1 MeV) and other particles as well as exposure to very high temperatures\cite{cook-2002-a}.

Ion implantation of semiconductors is well understood\cite{williams-1998-a,boer-2002-a}.
The type of scattering experienced by the implanted ion depends on
the mass of the ion and the relative angle of its trajectory to the
crystallographic axis. The distribution of implanted ions is a gaussian
centred about the projected range. The projected range in turn is
determined by the stopping power which has contributions both from
the nuclei in the crystal and from the electrons. The stopping power
($S$) is defined by the the projected range ($R_{p}$) through $R_{p}=\int_{0}^{E_{c}}{\rm d}E/S(E)$,
and is a sum of the nuclear and electronic components ($S=S_{n}+S_{e}$).
The electronic component has a very simple form, namely $S_{e}(E)=k_{e}\sqrt{E}$,
where $k_{e}$ is a constant (in silicon it has the value $10^{7}\sqrt{{\rm eV}}$/cm).
As the stopping power has the dimensions of a force, and is proportional
to the velocity of the incoming ion ($v=\sqrt{2E/M}$), it is behaving
as a kind of friction.

There has been considerable interest in the properties of candidate
materials for fusion powerplants under extreme conditions, and extensive
modelling work has been performed\cite{bacon-2002-a}. Clearly the
collisions between atoms to form cascades is an extremely important
process, but the dynamics of the atoms is significantly modified by
the response of the electrons to this violent motion. The dynamics
of radiation damage is typically divided into three phases: the displacement
phase, the relaxation phase and the cooling phase. The details of
the interaction between the nuclei and electrons need not be the same
in each phase because of the large variation in the energy of both
the nuclei and the electrons between the phases. Further, the manner
in which electrons transport heat is also modified when they become
highly excited\cite{huttner-1998-a}, or the lattice highly disordered\cite{raimondi-2004-a}.
There is still some controversy about precisely what happens, but
there are some general ideas which are clear. Highly energetic nuclei
can give up substantial amounts of energy to the electrons\cite{usman-2000-a},
and in so doing experience an effective friction force\cite{finnis-1991-a}.
As the coefficient of friction depends on materials parameters such
as electron-phonon coupling strength and electron heat capacity, it
leads to different rates of cooling in different materials, which
in turn can modify both defect formation and healing of damage\cite{finnis-1991-a}. 

There is a rather different type of process in which electrons (light
particles) are used to bombard a metal to produce damage (displacement
of heavy particles) though at quite a low level ($2\times10^{-6}$
to $2\times10^{-4}$ displacements per atom). Subsequent measurement
of the variation of resistance as a function of increasing temperature
during annealing of the damaged material provides direct information
about the diffusion barrier heights of point defects, which would
be difficult to obtain any other way\cite{fu-2005-a}.

\section{Theoretical methods and their applications }

As all theories involve some kind of simplification of reality, they
must all throw away features that are considered unimportant. However,
there will always be cases where what is normally unimportant suddenly
takes centre stage. When thinking about the interaction of atoms with
each other, it is usually a very good approximation to assume that
the electrons are so light, and thus move so fast relative to the
nuclei, that over the time it takes for the nuclei to undergo a significant
displacement (say $10^{-13}$s), the electrons will have undergone
so many collisions, both with nuclei (possibly as much as 1000 times
in a simple classical view) and with each other, that they will have
settled down into a well-defined state (normally taken to be their
ground state, or more generally that state which minimizes the chosen
free energy). Thus, we can treat the state of the electrons as known
once the \emph{positions} of the nuclei are given. This is the Born-Oppenheimer
approximation, whose chief characteristic is that it allows us to
treat the electrons and ions separately, and is ubiquitous because
it greatly simplifies the process of understanding matter at the atomic
scale.

However, the Born-Oppenheimer separation is an approximation, and
there are phenomena it cannot describe, notably those that are the
subject of this review. To understand the nature of this approximation,
and why it is unable to describe the irreversible transfer of energy
between electrons and nuclei, we need only note the following. Even
if the nuclei start at rest, collisions with electrons will result
in small energy transfers to nuclei (of order 0.1\% or less of the
electron energy per collision, classically). Once they have started
to fluctuate, the nuclei can transfer kinetic energy back to the electrons
in individual electron-nuclear collisions. But this two-way inelastic
energy transfer depends on the \emph{velocities} of the colliding
particles. This dependence lies beyond the Born-Oppenheimer approximation.

Fortunately, the amount of energy transferred in one collision is
very small. Thus the departure from the Born-Oppenheimer approximation
is a rather weak effect and so it makes sense to use perturbation
theory. The perturbation can be characterised by those contributions
to the forces on the nuclei that involve \emph{two} electronic states,
corresponding to electrons scattering from one state to another. The
unperturbed electronic states are some representation of the Born-Oppenheimer
energy surfaces, and for the nuclei harmonic oscillator states associated
with small displacements from equilibrium positions on those surfaces
are usually used (see \ref{sec:The-electron-phonon-Hamiltonian}).
Lowest order perturbation theory can then be used to evaluate the
rate of change of some quantity (such as the degree of excitation
of the nuclear vibrations) as a result of the perturbation. For systems
in which well-defined reference states can be defined, this prescription
provides not only insight into mechanisms, but also numbers that can
be compared with experiment. Examples include Joule heating in metals
and non-radiative transitions at point defects in semiconductors.

However, there are systems in which it is difficult to set up the
reference systems. If the nuclei undergo substantial displacements,
so that the concept of a reference position is not well-defined (for
example, flexible polymer strands in contact with a heat bath, or
a metal subject to high-energy bombardment) then neither the electronic
nor nuclear states are straightforward to characterise. However, such
systems have the feature that much of their behaviour can be well
described by molecular dynamics carried out within the Born-Oppenheimer
approximation. So, the natural consequence is to see if the standard
molecular dynamics algorithms can be revised to include the effects
of the breakdown of the Born-Oppenheimer approximation. Since the
full coupled problem is insoluble except in the simplest of cases
we have to investigate possible approximations.

In the following sections we discuss possible mathematical methods,
both perturbative and molecular dynamics based, that can be used to
investigate non-adiabatic processes. We begin with a simple classical
model in which both the nuclei and the electrons are treated as classical
particles obeying Newton's equations of motion. This cannot be considered
a particularly realistic model, but it is simple to understand and
in fact reveals much of the physics. We then proceed with the methods
that make explicit use of Born-Oppenheimer surfaces and oscillator
states, both perturbative and non-perturbative. Finally we survey
the methods based on molecular dynamics.

\subsection{Classical model }

While a classical model of the motion of electrons may not provide
us with a quantitative scheme in general, it makes it easy to understand
inelastic processes. To illustrate the points made we will look at
the heating of an atom by a current of electrons\cite{horsfield-2004-a}
.

The physical setup is as follows. We have an atom in a solid for which
the influence of the neighbouring atoms is represented by a spring
(an Einstein model). The electrons we treat as independent particles
which can interact with the oscillating atom. The Hamiltonian for
the system is the sum of an electron Hamiltonian ($H_{e}$), an atom
Hamiltonian ($H_{A}$) and a coupling Hamiltonian ($H_{eA}$)\begin{equation}
H=\underbrace{\sum_{i}\left(\frac{p_{i}^{2}}{2m}+v(\vec{r}_{i})\right)}_{H_{e}}+\underbrace{\left(\frac{P^{2}}{2M}+\frac{1}{2}KX^{2}\right)}_{H_{A}}-\underbrace{\sum_{i}\vec{X}\cdot\vec{\nabla}v(\vec{r}_{i})}_{H_{eA}}\label{eq:CM01}\end{equation}
where $v(\vec{r}_{i}-\vec{X})$ is the interaction potential between
electron $i$ and the atom, and we have made the approximation $v(\vec{r}_{i}-\vec{X})\approx v(\vec{r}_{i})-\sum_{i}\vec{X}\cdot\vec{\nabla}v(\vec{r}_{i})$,
which corresponds to small atomic displacements. Using Hamilton's
equations we can write down the equations of motion for the electrons
and for the atom\begin{eqnarray}
M\ddot{X}_{\nu} & = & -KX_{\nu}+\sum_{i}\frac{\partial v(\vec{r}_{i})}{\partial r_{i\nu}}\nonumber \\
m\ddot{r}_{i\nu} & = & -\frac{\partial v(\vec{r}_{i})}{\partial r_{i\nu}}+\sum_{\nu'}X_{\nu'}\frac{\partial^{2}v(\vec{r}_{i})}{\partial r_{i\nu'}\partial r_{i\nu}}\label{eq:CM02}\end{eqnarray}
In the absence of the coupling Hamiltonian $H_{eA}$, from Eq. \ref{eq:CM02}
we can write down an explicit solution for the atomic motion ($X_{\nu}(t)=A_{\nu}\sin(\sqrt{K/M}t+\phi_{\nu})$).
Furthermore, the electrons move independently of one another, with
each electron moving with a constant energy (the scattering from the
atom is elastic). Once we include the coupling Hamiltonian, this all
changes. We can no longer write down a simple closed form expression
for the atomic motion as it now depends on all the electrons; the
electrons are no longer completely independent because their motion
depends on $\vec{X}$, which in turn depends on all the electrons;
and electrons no longer move with constant energy because they experience
an external time varying force. It is the third point that is the
main topic of this review.

We can solve these equations for the case of motion in one dimension
in which the atom appears as a hard wall potential to the electrons.
In this case, by conserving energy and momentum during a collision,
we get the following approximate expression\cite{horsfield-2004-a}
for the power delivered to the atom ($w$) by a current ($j$) of
electrons in the limit that the electrons are much lighter than the
atom\begin{equation}
w\approx4\, j\,\left(\frac{m}{M}\right)\,(K_{e}-2K_{A})\label{eq:CM03}\end{equation}
where $K_{e}$and $K_{A}$ are the mean kinetic energies for the electrons
and atom respectively. We see that there are two contributions, one
positive from the electrons (which provides heating) and one negative
from the atom (which leads to cooling). The energy scale relevant
to the electron kinetic energy is $eV$, where $V$ is the applied
bias. At low temperatures, the energy scale relevant to the atom kinetic
energy is $\hbar\omega$ where $\omega$ is the vibrational frequency
of the atom. Thus for very small bias we might expect cooling, but
for typical voltages we expect the electrons to heat the atom, as
is indeed observed.

Equations \ref{eq:CM02} enable us also to see explicitly the microscopic
correlated electron-nuclear fluctuations that nucleate non-radiative
inelastic processes and mediate the inelastic exchange of energy between
the two subsystems. Consider an electron bound in an orbit with radius
$r$ and speed $v$ around a nucleus. The nucleus starts off at rest,
at the bottom of the parabolic well provided by the rest of the lattice.
Let us solve Eq. \ref{eq:CM02} approximately as follows. To lowest
order, ignore the motion of $X$ in the second equation. Then the
electron remains in the original orbit and provides a sinusoidally
varying external force on the nucleus, of amplitude $F=mv^{2}/r$
and frequency $\omega=v/r$, along each of the two axes in whose plane
the orbit lies. The nucleus in turn is now a driven harmonic oscillator.
If we solve the equation motion for the nucleus assuming that it starts
at rest, then the average kinetic energy over one period of oscillation
of the nucleus is given by\[
K_{A}=\frac{1}{2}\frac{F^{2}}{M}\frac{1}{2}\frac{3\omega^{2}+\omega_{0}^{2}}{(\omega^{2}-\omega_{0}^{2})^{2}}\]
 If $\omega\rightarrow\omega_{0}$, we see a resonance, with violent
heating of the nucleus. This resonance is the classical analogue of
the quantum transitions that become activated when the two frequencies
match. In the limit $\omega\gg\omega_{0}$, on the other hand, $K_{A}$
settles at \[
K_{A}=\frac{3}{2}\frac{m}{M}K_{e}\]
 where $K_{e}=mv^{2}/2$. The nucleus has acquired some kinetic energy,
and thus is no longer sitting still. The underlying motion of the
nucleus, responsible for $K_{A}$ above, is the wobble of a heavy
particle induced by the passage of a light particle, familiar from
planetary physics. As a consequence of this correlated wobble, nuclei
are never truly frozen: momentum conservation, with the consequence
that light and heavy particles always orbit a common centre of mass,
leads to a repartitioning of energy between the two systems, in a
ratio controlled by $m/M$.

\subsection{Born-Oppenheimer based electron-phonon coupling formalism}

In this section we consider those methods that make explicit use of
energy surfaces, and the non-adiabatic coupling between them.

\subsubsection{\label{sub:Landau-Zener-theory}Landau-Zener theory}

\begin{figure}
\begin{center}\includegraphics[%
  width=0.8\columnwidth,
  keepaspectratio]{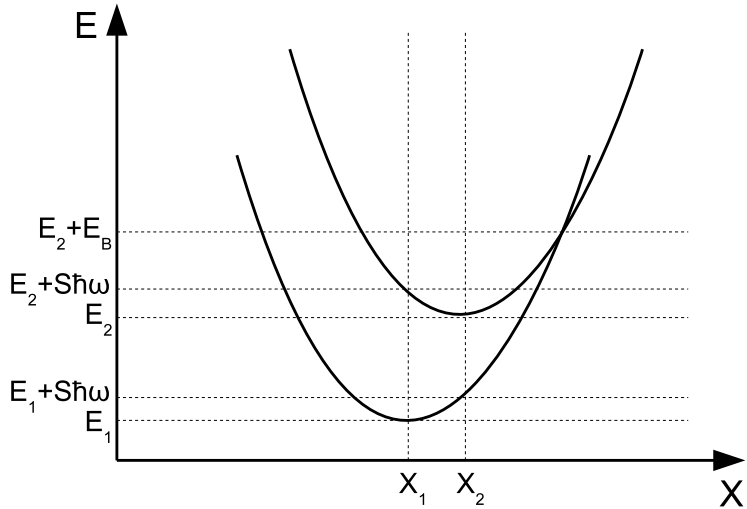}\end{center}

\caption{\label{fig:tls01}The parameters characerising a two level system.
The Landau-Zener theory concentrates on the crossing region on the
right-hand side of the diagram.}
\end{figure}

There is a generic model that has been successfully applied to a huge
number of problems: it is the two level harmonic model (or spin-boson
Hamiltonian) and is often considered in the presence of a dissipative
bath\cite{leggett-1987-a}. It has the advantage of providing useful
insights pictorially as well as through numbers. In figure \ref{fig:tls01}
is shown a representative system. The two diabatic surfaces (see \ref{sec:The-electron-phonon-Hamiltonian})
have the same curvature (vibrational frequency), but their minima
are offset relative to each other both in energy and position. There
are three other characteristic energies, two associated with vertical
(for example, optical) transitions ($E_{1}+S\hbar\omega$ and $E_{2}+S\hbar\omega$,
where $S$ is the Huang-Rhys factor), and one associated with non-radiative
transitions ($E_{2}+E_{B}$). It is the latter that interests us here.

The quantity $E_{B}$ is the barrier height that a classical oscillator
on the upper energy surface has to overcome to cross onto the lower
energy surface. The transition between the surfaces is a quantum process
(involving an electronic transition). This can correspond to a number
of physical phenomena, with an important one in the solid state being
electron capture (or emission) by a charge trap in a semiconductor\cite{henry-1977-a},
and another being atomic collisions with crystal surfaces\cite{tully-1977-a}.
This combination of quantum and classical descriptions was turned
into a transition rate in 1932 both by Zener and Landau\cite{zener-1932-a,landau-1932-a},
and has been extended in a number of ways since (see for example \cite{zhu-1994-a,kayanuma-1998-a}).

The essence of the original Landau-Zener theory is as follows. The
oscillating atoms (represented by just one coordinate, $X$ in figure
\ref{fig:tls01}, and from now on referred to as the oscillator) move
on a diabatic energy surface. The coupling to the neighbouring surface
is so weak (because of charge rearrangment in the case Zener studied)
that a transition is unlikely, so we can neglect the influence of
the transition on the dynamics of the oscillator for the purpose of
determining the transition rate. No transition is possible except
very near to the crossing. In this region the energy surfaces are
treated as linear, and the velocity of the oscillator as constant.
By solving the time dependent Schr\"odinger equation for the electrons
subject to the time varying potential due to the oscillator, the probability
of a transition during one pass of the oscillator over the crossing
is found to be\cite{zener-1932-a} $P_{X}=\exp(-2\pi\gamma)$ with
\begin{equation}
\gamma=\frac{|E_{12}|^{2}}{\hbar}\left(\frac{{\rm d}(E_{1}-E_{2})}{{\rm d}t}\right)^{-1}\label{eq:lz01}\end{equation}
where $E_{12}$ is the electronic matrix element coupling the two
surfaces, and $E_{1}-E_{2}$ is the energy difference between the
surfaces. The transition rate is then given by $\Gamma=2fP_{X}$\cite{ridley-1999-a}
where $f$ is the vibrational frequency.

\subsubsection{\label{sub:Perturbation-theory}Perturbation theory}

There is a long history of using perturbation theory for studying
non-adiabatic processes, both in the solid state and elsewhere. A
very important context for its use is in the Boltzmann equation treatment
of the transport properties of solids\cite{ziman-1960-a}. Fermi's
Golden Rule gives the transition rate from state $|i\rangle$ to state
$|f\rangle$\begin{equation}
\Gamma_{i\to f}=\frac{2\pi}{\hbar}\left|\langle f|\hat{H}_{1}|i\rangle\right|^{2}\delta(E_{f}-E_{i})\label{eq:PT01}\end{equation}
where $E_{i}=\langle i|\hat{H}_{0}|i\rangle$ and $E_{f}=\langle f|\hat{H}_{0}|f\rangle$,
$\hat{H}_{0}$ is the unperturbed Hamiltonian, and $\hat{H}_{1}$
is the perturbing Hamiltonian driving the transition. For the case
of interest to us (electrons interacting with mobile nuclei) from
\ref{sec:The-electron-phonon-Hamiltonian} we can identify $|i\rangle$
with $|\Psi_{nN}\rangle$, $|f\rangle$ with $|\Psi_{n'N'}\rangle$,
$E_{i}$ with $U_{nN}$ and $E_{f}$ with $U_{n'N'}$, where $n$
indexes electronic states and $N$ indexes nuclear states. To make
the integrals in the energies and transition matrix elements tractable,
the harmonic approximation is used for the nuclear states. For the
adiabatic representation (see \ref{sec:The-electron-phonon-Hamiltonian})
some further decisions about the dependence of the electronic wave
functions on nuclear coordinate is also necessary. These decisions
are implicit in the static lattice representation. In the context
of non-radiative transitions the choice of approximation is discussed
in detail by Stoneham in review articles \cite{stoneham-1981-a,stoneham-1990-a}
and a book\cite{stoneham-2001-a}. Once the rates are known, experimentally
observable quantities can be determined such as vibrational lifetimes
of adsorbed molecules\cite{persson-1980-a}, or the power supplied
to atoms ($w$) during electric current induced heating in nanocontacts\cite{todorov-1998-a,montgomery-2002-a}.
The power is given by\begin{equation}
\fl w=\frac{2\pi}{\hbar}\sum_{n'N'}\,|\langle\Psi_{n'N'}|\hat{H}_{1}|\Psi_{nN}\rangle|^{2}\,(W_{N'}-W_{N})\,\delta(U_{n'N'}-U_{nN})\label{eq:PT02}\end{equation}
where $W_{N}$ is the energy of the nuclei in state $N$. These matrix
elements have been calculated, within tight binding and density functional
theory, for atomic and molecular wires\cite{montgomery-2002-a,montgomery-2003-b,montgomery-2003-a,YChen03,YChen04,YChen05,yang-2005-a}.
The power into a phonon mode has the structure\begin{equation}
\fl w=\frac{\pi\hbar}{M}\sum_{\alpha\beta}\, f_{\alpha}(1-f_{\beta})\,|\langle\beta|\vec{g}|\alpha\rangle|^{2}\,\{(N+1)\delta[\epsilon_{\alpha}-(\epsilon_{\beta}+\hbar\omega)]-N\delta[\epsilon_{\alpha}-(\epsilon_{\beta}-\hbar\omega)]\}\label{eq:PT03}\end{equation}
where $|\alpha\rangle$ is a single particle electronic orbital with
eigenvalue $\epsilon_{\alpha}$ and occupancy $f_{\alpha}$, and $\vec{g}$
is the gradient of the electronic Hamiltonian with respect to the
oscillator normal coordinate. The phonon frequency is $\omega$ and
its occupancy is $N$. For simplicity we have assumed that all comic
masses are the same, $M$, but in practice this assumption is easily
lifted. Equation \ref{eq:PT03} accounts for the observed heating
in nanocontacts\cite{todorov-1998-a,montgomery-2002-a,YChen03}. The
same perturbative model can also account for some of the inelastic
current-voltage features in atomic and molecular wires\cite{montgomery-2003-b,YChen04,YChen05,yang-2005-a}.

Lowest order perturbation theory clearly cannot handle multiple coherent
transitions or strong interactions\cite{zhitenev-2002-a}. However,
higher order theories can be used in these cases (see \ref{sub:Non-equilibrium-Green's-functions}).
Problems where higher order theories have been used include the non-adiabatic
transition at a crossing between two electronic levels in the presence
of dissipation\cite{kayanuma-1998-a}, cross-sections for inelastic
scattering of electrons from a surface\cite{persson-1984-a} and inelastic
tunneling cross-sections\cite{persson-1986-a}.

\subsubsection{\label{sub:Non-equilibrium-Green's-functions}Non-equilibrium Green's
functions}

Green's functions are one of the well established tools for dealing
with problems in condensed matter. They provide information about
the response at any point in space-time due to an excitation (usually
the creation or annihilation of a particle) at any other point. Green's
functions are used in perturbation theory, linear response theory,
and quantum kinetic equations. Their real power becomes evident when
we consider a system of many particles (electrons and nuclei) that
interact, and more especially when the system is being driven out
of equilibrium by some external forces. \textit{\emph{Their application
to transport, in the framework of the Keldysh formalism,}} has a long
history\cite{caroli-1971-a,caroli-1971-b,combesco-1971-a,caroli-1972-a,martin-rodero-1988-a,ferrer-1988-a,YMeir92}.
In the absence of particle interactions, the many-body Keldysh formalism
reduces \cite{brandbyge-2002-a} to the steady-state transport formalism
based on one-electron Lippmann-Schwinger scattering theory (for a
brief review see \cite{todorov-2002-a}), which in turn reduces \cite{YMeir92}
to the Landauer formalism. In the following, we present some applications
of the Green's functions technique for systems with interactions between
electrons and atomic motion. Note that in this section (and the corresponding
appendices) we depart from the usual notation for representing all
operators by means of hats (such as $\hat{H}_{e}$) to avoid a confusion
of extra symbols. Instead hats are used to represent only operators
in the interaction picture.

\textit{Beyond perturbation theory:} Let us start by generalizing
Fermi's golden rule to include higher-order terms in the perturbation
$V$. The formal derivation based on scattering theory is rather involved,
therefore we consider a simpler approach following the prescription
given in Refs.~\cite{GDoyen93,HNess97,HBruus2004}. This will permit
us to introduce the different quantities that are needed in the full
description based on non-equilibrium statistical physics \cite{LKeldysh65,LKadanoff1962}.

Let us assume that the perturbation $V$ in the full Hamiltonian $H=H_{0}+V$
is turned on adiabatically in the distant past, \emph{i.e.} $V\equiv V{\rm e}^{\eta t/\hbar}$
(with $\eta$ a small positive constant). Initially the system is,
at time $t_{0}$, in the state $|i\rangle$ and evolves at time $t>t_{0}$
into the state $|i(t)\rangle$: \begin{equation}
|i(t)\rangle={\rm e}^{-{\rm i}H_{0}t/\hbar}\hat{U}(t,t_{0})\,\,{\rm e}^{{\rm i}H_{0}t_{0}/\hbar}\,|i\rangle\,,\label{i_oft}\end{equation}
 where $\hat{U}(t,t_{0})$ is the time evolution operator (given within
the interaction picture) of the system. The probability $P(t)$ to
find the system in the final state $|f\rangle$ at time $t$ is given
by $|\langle f|i(t)\rangle|^{2}$ and its time derivative gives the
transition rate $\Gamma_{fi}=\Gamma_{i\rightarrow f}$. To derive
the generalized Fermi's golden rule, we include all orders of the
perturbation $V$ in the time evolution operator $\hat{U}$. Then,
we have \begin{equation}
\langle f|i(t)\rangle=\langle f|\hat{U}(t,t_{0})|i\rangle\,{\rm e}^{-{\rm i}E_{f}t/\hbar}{\rm e}^{{\rm i}E_{i}t_{0}/\hbar}\,,\label{fi_oft}\end{equation}
 with\begin{eqnarray}
\fl\hat{U}(t,t_{0}) & = & \sum_{n}^{\infty}\frac{1}{({\rm i}\hbar)^{n}}\int_{t_{0}}^{t}{\rm d}t_{1}\int_{t_{0}}^{t_{1}}{\rm d}t_{2}\int_{t_{0}}^{t_{2}}{\rm d}t_{3}\dots\int_{t_{0}}^{t_{n-1}}{\rm d}t_{n}\hat{V}(t_{1})\hat{V}(t_{2})\dots\hat{V}(t_{n})\,{\rm e}^{\eta t_{n}/\hbar}\nonumber \\
\fl & = & \sum_{n=0}^{\infty}\frac{1}{n!}\frac{1}{({\rm i}\hbar)^{n}}\int_{t_{0}}^{t}{\rm d}t_{1}\dots\int_{t_{0}}^{t}{\rm d}t_{n}T_{t}\left(\hat{V}(t_{1})\dots\hat{V}(t_{n})\right)=T_{t}\left(\exp\left(-\frac{{\rm i}}{\hbar}\int_{t_{0}}^{t}{\rm d}\tau\hat{V}(\tau)\right)\right)\,,\label{Uttp}\end{eqnarray}
 where the perturbation $\hat{V}(t)$ is taken in the interaction
representation $\hat{V}(t)={\rm e}^{{\rm i}H_{0}t/\hbar}V\,{\rm e}^{-{\rm i}H_{0}t/\hbar}$
and $T_{t}$ denotes the time ordering operator. Performing the time
integrations and making use of the interaction picture for the perturbation,
we obtain \begin{equation}
P(t)=|\langle f|i(t)\rangle|^{2}=\left|\frac{{\rm e}^{\eta t/\hbar}}{E_{i}-E_{f}+{\rm i}\eta}\langle f|T|i\rangle\right|^{2}\,,\label{Poft}\end{equation}
 and identifying the time derivative of $P(t)$ with the transition
rate $\Gamma_{fi}$, one finds the generalised Fermi's golden rule:
\begin{equation}
\Gamma_{fi}=\frac{2\pi}{\hbar}|\langle f|T|i\rangle|^{2}\,\,\delta(E_{f}-E_{i})\,\,.\label{genFGR}\end{equation}
 The quantity $T$ in Eq.~(\ref{Poft}) is called the $T$-matrix
and is given by the series expansion \begin{equation}
\fl T=V+V\frac{1}{E_{i}-H_{0}+{\rm i}\eta}V+V\frac{1}{E_{i}-H_{0}+{\rm i}\eta}V\frac{1}{E_{i}-H_{0}+{\rm i}\eta}V+\dots\label{Tmatrix_series}\end{equation}
 For the lowest-order expansion $T\approx V$ or equivalently by taking
$\hat{U}(t,t_{0})\approx1-{\rm i}/\hbar\int_{t_{0}}^{t}{\rm d}t'\hat{V}(t')$
in Eq. (\ref{Uttp}), one recovers the transition rate $\Gamma_{fi}=\Gamma_{i\rightarrow f}$
given by the usual Fermi's golden rule Eq. \textbf{\ref{eq:PT01}}.

The $T$-matrix can be rewritten in a compact and closed form \begin{equation}
T=V+V\frac{1}{E_{i}-H_{0}+{\rm i}\eta}\,\, T=V+VG_{0}^{r}\,\, T\label{Tmatrix_series2}\end{equation}
 which generates the same series as in Eq.~\ref{Tmatrix_series}.
This allows us to introduce one of the different Green's functions,
namely the retarded Green's function $G_{0}^{r}$ defined, in the
energy representation, by $G_{0}^{r}(\omega)=[\omega-H_{0}+{\rm i}\eta]^{-1}$.
This Green's function is obtained from the unperturbed Hamiltonian
$H_{0}$. The retarded Green's function $G^{r}$ for the total Hamiltonian
$H=H_{0}+V$ is defined by $G^{r}(\omega)=[\omega-H+{\rm i}\eta]^{-1}$.
Then, the $T$-matrix can be expressed as $T=V+VG_{0}^{r}\,\, T=V+VGV$
by making use of the Dyson equation that links the full Green's function
to the unperturbed Green's function: $G^{r}=G_{0}^{r}+G_{0}^{r}VG^{r}=G_{0}^{r}+G_{0}^{r}TG_{0}^{r}$.
Now we have all the ingredients to develop the non-equilibrium theory,
namely the concept of Green's functions, the time evolution operator
and the Dyson equation. One already sees the importance of using Green's
functions in determining transition rates when going beyond perturbation
theory. Obviously, the conventional results of perturbation theory
are obtained by expanding the Green's function in the Dyson equation
or the $T$-matrix to the lowest-order in the perturbation.

The $T$-matrix formalism with scattering boundary conditions has
been used to study the effects of electron-atomic vibration coupling
in the transport properties of atomic and molecular wires. This was
done \textit{\emph{(i)}} to the lowest-order in the interaction, \emph{i.e.}
perturbation theory, in atomic wires \cite{montgomery-2003-b,montgomery-2002-a,montgomery-2003-a}
and molecular wires \cite{YChen05,YChen04,YChen03}; and \textit{\emph{(ii)}}
to all orders in the interaction in model systems and molecular wires
by using conventional Green's functions techniques \cite{FSols92,JBonca95,HNess99,KHaule99,NMingo00,HNess01,HNess02,HNess02b,ATroisi03,ATroisi05,JJiang05,bihary-2005-a}.
The connection between the inelastic $T$-matrix scattering formalism
and non-equilibrium Green's functions has been discussed in Ref.~\cite{HNess05}.\\

\textit{Interaction as self-energy in the Green's functions:} To describe
the ground state or the thermal-averaged properties of a system of
many particles, such as the coupled electron/atom system considered
in this paper, it is useful to work with a many-body approach based
on Green's functions. Such a Green's functions approach is even more
useful and powerful when we want to calculate the properties of the
system driven out of equilibrium by some external forces. The exact
definitions of the different Green's functions and their interrelation
is given in \ref{sec:Equilibrium-and-non-equilibrium} as well as
the principles used to derive the theory for non-equilibrium conditions.

Now, it is important to define precisely what the perturbation $V$
means physically. Here we consider $V$ as being a perturbation on
the reference system (described by the unperturbed Hamiltonian $H_{0}$)
due to some kind of interaction. It is convenient to distinguish between
different kinds of interaction, though generally speaking the calculation
of such interactions and their inclusion in the Green's functions
are formally equivalent for all kinds of interactions. In the following,
it is convenient to consider as separate the following interactions: 

\begin{enumerate}
\item The interactions between particles, \emph{i.e.} the interactions between
electrons, between phonons, or the interaction between electrons and
phonons (or atomic motion). In practice, such interactions are incorporated
in the corresponding electron or phonon Green's functions under the
form of a proper self-energy via a Dyson-like equation for the Green's
function. In the literature, it is usual to talk about electron Green's
functions being dressed by the phonons, and phonon Green's functions
being dressed by the electrons. The corresponding self-energies are
usually difficult to calculate exactly for a many-particle system.
They are obtained, up to some order in the interaction parameters,
via the use of a many-body Feynman diagrammatic perturbation theory
\cite{GMahan1990,JRammer86}.
\item The interactions of the particles with an external field. As an example,
one can consider an electromagnetic field exciting the electronic
system by inducing electronic transitions, or an external applied
bias that drives a nanojunction out of equilibrium by establishing
a steady-state current flow. The corresponding hot electrons would
then transfer energy to the phonon degrees of freedom via some interaction
of type (i) mentioned above.
\item It is often interesting to partition a system into its different constituent
parts. For example: a crystal partioned into a localized region around
a defect and the rest of the crystal; a surface with adsorbate partitioned
into the bare surface and the adsorbate; a nano-junction partitioned
into three parts (a central region whose transport properties are
studied and the two electrodes to which the central region is connected).
In this case, the interaction between two different parts (I) and
(II) of the system (for example the part of the electronic Hamiltonian
that couples regions (I) and (II)) appears under the form of a self-energy
in the Green's functions expanded onto the subspace of one of these
regions. Often these self-energies are also referred to as embedding
potentials \cite{JInglesfield81}. These self-energies are usually
practical to calculate, especially for systems treated within a mean-field
approach and when there is no crossing of the interactions of type
(i) between regions (I) and (II). 
\end{enumerate}
We now consider that the system has either reached a stationary-state
regime or more simply is at equilibrium. Then the Green's functions
depend only on the time difference $t-t'$, and their Fourier-transform
$G^{x}(\omega)$ depend on only one energy argument $\omega$. In
the presence of interactions, the electron Green's function $G^{r,a}$
obeys a Dyson equation (either at equilibrium or in a non-equilibrium
regime): $G^{r,a}(\omega)=G_{0}^{r,a}(\omega)+G_{0}^{r,a}(\omega)\Sigma^{r,a}(\omega)G^{r,a}(\omega)$
where $G_{0}^{r,a}$ is the Green's function of the electronic system
in the absence of the interactions, and $\Sigma^{r,a}$ is the self-energy
arising from these interactions. A similar Dyson equation relates
the phonon Green's function $D^{r,a}$ in the presence of interactions
to the undressed phonon Green's function $D_{0}^{r,a}$ (see \ref{sec:Equilibrium-and-non-equilibrium}).
For the non-equilibrium state, the Green's functions $G^{<,>}$ and
$D^{<,>}$ obey another kind of quantum kinetic equation (see below
or \ref{sec:Equilibrium-and-non-equilibrium}).

The difficulty is to calculate exactly, or as accurately as possible,
the different Green's functions by including all orders of the different
kinds of interactions. It is also a challenge to calculate them numerically
because the self-energies $\Sigma^{x=(r,a,<,>)}$ arising from the
interactions are actually functionals of the different Green's functions
themselves: $\Sigma^{x}(\omega)=\Sigma^{x}[\{ G^{x}(\omega)\},\{ G_{0}^{x}(\omega)\}]$
where $G^{x}$ ($G_{0}^{x}$) denotes a Green's function in the presence
(absence) of the interactions respectively.

One way of solving the problem approximately is to expand the Dyson
equation in a Born series of the Green's function $G_{0}$ in the
absence of the interactions, $G=G_{0}+G_{0}\Sigma G_{0}+G_{0}\Sigma G_{0}\Sigma G_{0}\Sigma G_{0}+...$,
then choose lower-order Feynman diagrams in the interaction parameters
for the proper self-energy $\Sigma$ (the Born approximation), and
finally substitute the Green's function $G_{0}$ in the expression
of the self-energy by the full Green's function $G$. In this way,
one introduces a self-consistent scheme since the Green's function
$G$ both determines and is determined by the proper self-energy $\Sigma$.
This approximation is usually known as the self-consistent Born approximation
(SCBA). Physically it means that the interaction is treated to all
orders but that only a limited number of elementary excitations are
generated, \emph{i.e.} those given by the many-body diagrams chosen
for the self-energy. Within the SCBA some processes involving crossed
diagrams are omitted (if they were not already included in the self-energy).

Now that we have described the principle of non-equilibrium Green's
functions and the ways to include the interactions, we consider an
example of their use in relation to electronic transport through a
heterojunction in the presence of an electron-phonon interaction.
Then we will briefly describe how they can be used to derive quantum
kinetic equations as a generalisation of the Boltzmann equation for
transport.\\

\textit{Electronic current in the presence of interaction:} Let us
consider a scattering central region (a quantum dot, a molecular or
atomic wire) in which there are interactions between particles, and
which is connected to two (left $L$ and right $R$) leads. The latter
are described by two non-interacting Fermi seas at their own equilibrium
and thus characterised by two Fermi distributions $f_{L}$ and $f_{R}$.
The electronic current $J_{L}$ flowing from the left lead into the
central region is then expressed in terms of three Green's functions
($G^{r,a}$ and $G^{<}$) of the interacting central region. In the
stationary-state regime, the current $J_{L}$ is given by \cite{caroli-1971-a,caroli-1971-b,caroli-1972-a,YMeir92,SDatta1995}:
\begin{equation}
\fl J_{L}=\frac{{\rm i}e}{h}\int\textrm{d}\omega\,\textrm{Tr}\left\{ f_{L}(\omega)\,\Gamma_{L}(\omega)\,[G^{r}(\omega)-G^{a}(\omega)]+\Gamma_{L}(\omega)\, G^{<}(\omega)\right\} \,,\label{Imeir1}\end{equation}
 where the trace runs over the electron states of the central region
and $\Gamma_{L}$ is related to the imaginary part of the retarded
(advanced) self-energy $\Sigma_{L}^{r(a)}$ arising from the coupling
of the central region to the left lead (self-energy arising from an
interaction of type (iii) --- see above). These self-energies and
other self-energies arising from the other kinds of interactions should
be included in the calculation of $G^{r(a)}$.

An expression similar to Eq.~\ref{Imeir1} can be obtained for the
current $J_{R}$ flowing from the right lead into the central region
by interchanging the subscript $L\leftrightarrow R$ in Eq.~\ref{Imeir1}.
For a current conserving system, one has $J_{L}=-J_{R}$. The famous
result of Meir and Wingreen \cite{YMeir92} is obtained from the symmetrised
current $J=\frac{1}{2}(J_{L}-J_{R})$\begin{equation}
\fl J=\frac{{\rm i}e}{2h}\int\textrm{d}\omega\,\textrm{Tr}\left\{ \left(f_{L}(\omega)\,\Gamma_{L}(\omega)-f_{R}(\omega)\,\Gamma_{R}(\omega)\right)[G^{r}(\omega)-G^{a}(\omega)]+\left(\Gamma_{L}(\omega)-\Gamma_{R}(\omega)\right)G^{<}(\omega)\right\} \label{eq:IMeirWingreen}\end{equation}

By using the relationship between the different Green's functions:
$G^{r}-G^{a}=G^{>}-G^{<}$ (\emph{c.f.} \ref{sec:Equilibrium-and-non-equilibrium}),
Eq.~\ref{Imeir1} can be rewritten as $J_{L}=e/h\int\textrm{d}\omega\,\textrm{Tr}\{\Sigma_{L}^{<}(\omega)\, G^{>}(\omega)-\Sigma_{L}^{>}(\omega)\, G^{<}(\omega)\}$,
where the self-energies due to the coupling of the central region
to the left lead are $\Sigma_{L}^{<}\propto{\rm i}f_{L}\Gamma_{L}$
and $\Sigma_{L}^{>}\propto-{\rm i}(1-f_{L})\Gamma_{L}$. The physical
interpretation of this equation is more transparent than Eq.~\ref{Imeir1}.
The first term gives the current flowing through the left contact
from the left electrode towards the central region, because it includes
$G^{>}$ which gives information about the empty non-equilibrium states
in the central region and the self-energy $\Sigma_{L}^{<}$ which
is proportional to the distribution of occupied states in the left
lead. The second term gives the current flowing through the left contact
towards the electrode because it includes $G^{<}$ which gives information
about the occupied non-equilibrium states of the central region, and
the self-energy $\Sigma_{L}^{>}$ is proportional to left lead empty
states.

Now, to get physical results for the current we need to calculate
the different Green's functions involved in Eq.~\ref{Imeir1}. These
functions obey Dyson-like equations $G^{r,a}=G_{0}^{r,a}+G_{0}^{r,a}\Sigma^{r,a}G^{r,a}$
for $G^{a,r}$ and another quantum kinetic equation for $G^{<,>}$,
\emph{i.e.} $G^{<,>}=(1+G^{r}\Sigma^{r})G_{0}^{<,>}(1+\Sigma^{a}G^{a})+G^{r}\Sigma^{<,>}G^{a}$.
The first term in the equation for $G^{<,>}$ comes from the initial
conditions and is seen as a boundary condition. It is generally omitted
for the stationary-state since it acts only in the transient regime.
The approach is now very useful if we can construct the self-energy
functionals that include the physics of the many-body system considered
and that a sufficiently good solution can be obtained from the coupled
equations for $G^{a,r}$ and $G^{<,>}$ (\emph{c.f.} \ref{sec:Equilibrium-and-non-equilibrium}).
One important point is that the self-energy functionals and the solution
obtained for the Green's functions preserve the condition of current
conservation. It can be shown that within the decomposition made above
for the different kinds of interaction, the condition for current
conservation requires that $\int{\rm d}\omega\,\textrm{Tr}\{\Sigma_{\textrm{inel}}^{<}(\omega)G^{>}(\omega)-\Sigma_{\textrm{inel}}^{>}(\omega)G^{<}(\omega)\}=0$,
where $\Sigma_{\textrm{inel}}^{<,>}$ are the self-energies due to
interactions of types (i) and (ii) as defined above. This condition
says that what is inelastically scattered into the central region
should compensate what is inelastically scattered out of the central
region. It is not obvious that all choices for the self-energy functionals
fullfil this condition. A good example is the SCBA for the electron-phonon
interaction calculated from the undressed phonon Green's function
$D_{0}$. The SCBA has been used recently to study the effects of
the coupling between electrons and atomic motion in atomic wires \cite{TFrederiksen04,MPaulsson05}.
Other self-energy functionals, developed in the spirit of the SCBA
and including more elaborate approximations for the phonon propagators,
have also been used to study the effects of the interaction between
electrons and atomic vibrations in model systems \cite{PHyldgaard94,TMii02,TMii03,MGalperin04b,AMitra04,DRyndyk05,galperin-2004-a}
and in more realistic atomic and molecular wires \cite{APecchia04,APecchia04b,YAsai04,YAsai04err,JViljas05,LVega05}.
However the problem of non-equilibrium transport through a coupled
electron-phonon system is very complex to solve exactly, and more
work towards this direction needs to be done.\\

\textit{Quantum kinetic equations:} It is possible to derive a quantum
analog to the Boltzmann semiclassical equation for transport from
the non-equilibrium Green's functions. The detailed derivation is
rather complex and has been well described in other review articles
\cite{JRammer86,GMahan87,JRammer91}. Here we summarise the principles
for obtaining such quantum kinetic equations.

We start with the lesser Green's function $G^{<}$ given in a space-time
representation $(r,t)$ for a fermion field operator in a solid. Then
we perform a transformation similar to that given in \ref{sec:The-Wigner-transform}
by defining the centre-of-mass and time-average coordinates $(R,T)\equiv\frac{1}{2}(r_{1}+r_{2},t_{1}+t_{2})$
and the relative coordinates $(x,\theta)\equiv(r_{1}-r_{2},t_{1}-t_{2})$.
The Green's function $G^{<}(r,\theta;R,T)$ expressed in these new
coordinates is Fourier transformed in space and time with respect
to the relative coordinates to give $G^{<}(k,\omega;R,T)$, from which
is defined the so-called Wigner distribution function $f(k,\omega;R,T)={\rm i}G^{<}(k,\omega;R,T)$.
The distribution $f(k,\omega;R,T)$ is the quantum analog of the semiclassical
distribution function $f(r,v,t)$ used in the Boltzmann equation for
transport ($v$ being the particle velocity at point $(r,t)$)%
\footnote{Note that the function $f(k,\omega;R,T)={\rm i}G^{<}(k,\omega;R,T)$
is related to the Wigner matrix used in CEID (see section \ref{sub:Correlated-electron-ion-dynamics})
by an integral over the energy $\omega$\cite{jacoboni-bordone-2004}.
However, as used in CEID the transform is over nuclear degrees of
freedom, while here it is over electronic ones.%
}. In a system governed by the laws of quantum mechanics, because of
the uncertainty principle and because scattering smears out the energy
and momentum states, the usual relation between energy and velocity
does not hold any more. Thus we have to consider a distribution function
of the type $f(k,\omega;R,T)$ where the energy $\omega$ and momentum
$k$ are treated as independent variables.

In order to get the quantum Boltzmann equation, we write the equation
of motion for the Green's function $G^{<}(k,\omega;R,T)$. This is
the most difficult part of the problem especially when we want to
treat a many-body system with interactions \cite{GMahan87,JRammer91}.
However the quantum distribution function $f(k,\omega;R,T)$ is the
only correct distribution function to describe particles in an interacting,
many-body approach. In the presence of an electric field, the following
quantum Boltzmann equation can be derived \cite{GMahan87}\begin{equation}
\frac{\partial}{\partial t}f+v\cdot\nabla_{r}f+F\cdot\left\{ \frac{\nabla_{k}}{m}+v\frac{\partial}{\partial\omega}\right\} f+\left(\frac{\partial}{\partial t}f\right)_{S}=0\,,\label{QKEorQBE}\end{equation}
 where $F$ is the force (electric field) acting on the electron.
An additional driving term ($v\cdot F\partial f/\partial\omega$)
is obtained in comparison to the usual Boltzmann equation. The last
term $(\partial f/\partial t)_{S}$ arises from the inelastic scattering
due to interaction between particles. It can be calculated by the
use of self-energies $\Sigma(\omega)$ (arising from electron-electron
or electron-phonon interactions) in the corresponding equation of
motion for $G^{<}(k,\omega;R,T)$. Once Eq.~\ref{QKEorQBE} is solved
for the distribution function $f(k,\omega;R,T)$, we can calculate
the macroscopic quantities, as measured in various experiments, such
as the electronic density $n(R,T)$, the electronic current $j(R,T)$,
the energy density $n_{E}(R,T)$ and energy current $j_{E}(R,T)$.
The latter are obtained from the different moments (in energy $\omega$
or momentum $k$) of the distribution function $f$. Finally, the
semiclassical Boltzmann distribution $f(R,v,T)$ is found by integrating
the distribution function $f(k=mv,\omega;R,T)$ over energy.

\subsection{Molecular dynamics based methods}

There have been many suggested ways to modify the standard Born-Oppenheimer
based molecular dynamics algorithms. It is usual to speak of their
being broad categories of methods: surface hopping and effective interaction.
The former category is very popular for studies of molecular systems,
but has well-known deficiencies (it is somewhat \emph{ad hoc}, and
coherence between trajectories is lost). The latter approach (usually
in the Ehrenfest approximation) is popular for some problems but also
has limitations. One recent method (Correlated Electron-Ion dynamics\cite{horsfield-2004-b,horsfield-2005-a,todorov-2005-a,bowler-2005-a})
sits somewhere between the two usual categories, though it is viewed
as an extension of the Ehrenfest approximation. We begin with a discussion
of the Ehrenfest method because of its importance as a starting point
for the other methods. We then present alternative methods that attempt
to overcome its deficiencies.

\subsubsection{Ehrenfest method}

This is a simple, but ingenious, method that successfully adds some
non-adiabatic features to molecular dynamics, but is incomplete. There
are many ways to derive the formal equations for this method, but
the one we shall use here most naturally leads us to CEID which is
discussed in detail below.

The Ehrenfest method consists of two separate approximations: the
electrons interact with the nuclei through a classical field generated
by its charge distribution (a mean field approximation, and the main
source of error); the nuclei obey classical (Newtonian) mechanics.
We will work with the density matrix.

The first approximation allows us to write the full density matrix
$\hat{\rho}$ as a tensor product of an electron density matrix $\hat{\rho}_{e}$
and a nuclear density matrix $\hat{\rho}_{N}$: $\hat{\rho}=\hat{\rho}_{e}\otimes\hat{\rho}_{N}$.
If we substitute this into the quantum Liouville equation\begin{equation}
\frac{\mathrm{d}\hat{\rho}}{\mathrm{d}t}=\frac{1}{\mathrm{i}\hbar}\left[\hat{H},\hat{\rho}\right]\label{eq:efest-01}\end{equation}
 and then trace out either the nuclear or electronic degrees of freedom
we obtain the following equations of motion:\begin{eqnarray}
\frac{\mathrm{d}\hat{\rho}_{e}}{\mathrm{d}t} & = & \frac{1}{\mathrm{i}\hbar}\left[\bar{H}_{e},\hat{\rho}_{e}\right]\nonumber \\
\frac{\mathrm{d}\hat{\rho}_{N}}{\mathrm{d}t} & = & \frac{1}{\mathrm{i}\hbar}\left[\bar{H}_{N},\hat{\rho}_{N}\right]\label{eq:efest00}\end{eqnarray}
where $\bar{H}_{e}=\mathrm{Tr}_{N}\left\{ \hat{H}_{e}\hat{\rho}_{N}\right\} $
, $\bar{H}_{N}=\hat{T}_{N}+\mathrm{Tr}_{e}\left\{ \hat{H}_{e}\hat{\rho}_{e}\right\} $,
$\mathrm{Tr}_{N}\left\{ \dots\right\} $ means a trace over nuclear
coordinates, and $\mathrm{Tr}_{e}\left\{ \dots\right\} $ means a
trace over electronic coordinates. The total Hamiltonian ($\hat{H}$)
has been separated into the nuclear kinetic energy ($\hat{T}_{N}$)
and the rest ($\hat{H}_{e}$). It is the presence of the traces in
the definitions of the effective Hamiltonians that results in the
electrons and nuclei only responding to the classical fields. Note
that below we show how the Ehrenfest approximation can lift the mean
field approximation for the nuclei because they can be treated as
classical particles with unique trajectories.

The classical nature of the nuclei is introduced by replacing the
quantum commutator brackets by classical Poisson brackets. The equation
of motion for the classical nuclear density matrix (now a phase space
density) becomes\begin{equation}
\frac{\mathrm{d}\rho_{N}(\vec{R},\vec{P})}{\mathrm{d}t}=\sum_{\nu}\left(\frac{\partial\bar{H}_{N}(\vec{R})}{\partial R_{\nu}}\frac{\partial\rho_{N}(\vec{R},\vec{P})}{\partial P_{\nu}}-\frac{P_{\nu}}{M_{\nu}}\frac{\partial\rho_{N}(\vec{R},\vec{P})}{\partial R_{\nu}}\right)\label{eq:efest01}\end{equation}
where $R_{\nu}$ and $P_{\nu}$ are the position and momentum coordinates
and $M_{\nu}$ is a nuclear mass. If we consider just a single classical
trajectory (indicated by the subscript $T$) then we have:\begin{equation}
\rho_{N}(\vec{R},\vec{P})=\prod_{\nu}\delta\left(R_{\nu}-R_{T,\nu}(t)\right)\delta\left(P_{\nu}-P_{T,\nu}(t)\right)\label{eq:efest02}\end{equation}
where $R_{T,\nu}(t)$ and $P_{T,\nu}(t)$ are the nuclear positions
and momenta along the trajectory. Substituting Eq. \ref{eq:efest02}
into Eq. \ref{eq:efest01} we obtain\begin{eqnarray}
\dot{R}_{T,\nu} & = & \frac{P_{T,\nu}}{M_{\nu}}\nonumber \\
\dot{P}_{T,\nu} & = & -\frac{\partial\bar{H}_{N}(\vec{R}_{T})}{\partial R_{T,\nu}}=-\mathrm{Tr}_{e}\left\{ \hat{\rho}_{e}\frac{\partial\hat{H}_{e}(\vec{R}_{T})}{\partial R_{T,\nu}}\right\} \label{eq:efest03}\end{eqnarray}
Inserting this nuclear density (Eq. \ref{eq:efest02}) into the equation
of motion for the electrons (Eq. \ref{eq:efest00}) we obtain\begin{equation}
\frac{\mathrm{d}\hat{\rho}_{e}}{\mathrm{d}t}=\frac{1}{\mathrm{i}\hbar}\left[\hat{H}_{e}(\vec{R}_{T}),\hat{\rho}_{e}\right]\label{eq:efest04}\end{equation}
Equations \ref{eq:efest03} and \ref{eq:efest04} constitute the Ehrenfest
approximation. The use of a \emph{single} nuclear trajectory in fact
takes us halfway to introducing microscopic fluctuations: the electrons
now respond to \emph{individual} nuclei. However, the converse is
not true, with the nuclei still responding to an average distribution
of electrons.

This approximation can fail either when the nuclei have to be treated
as quantum particles (for example, when tunnelling takes place), or
the nuclei respond to the microscopic fluctuations in the electron
charge density as well as the mean (for example, Joule heating). In
this review we are principally concerned with the latter case.

To make this more transparent, let us suppose that the electronic
density matrix can be represented by a phase space distribution $\rho_{e}(\vec{r},\vec{p})$
(this can be formally justified by means of the Wigner transform as
described in \ref{sec:The-Wigner-transform}). In this case we have
$\bar{H}_{N}(\vec{R})=\int\mathrm{d}\vec{r}\,\mathrm{d}\vec{p}\,\rho_{e}(\vec{r},\vec{p})H_{e}(\vec{R};\vec{r},\vec{p})$,
and hence the force on the nuclei ($\bar{F}_{\nu}$) due to the electrons
and the other nuclei is (Eq. \ref{eq:efest03}) $\bar{F}_{\nu}=-\int\mathrm{d}\vec{r}\,\mathrm{d}\vec{p}\,\rho_{e}(\vec{r},\vec{p})\partial H_{e}(\vec{R};\vec{r},\vec{p})/\partial R_{\nu}$.
If, as for the nuclei, we could isolate a \emph{single} electronic
trajectory $\vec{r}_{T}(t)$, then the force would just be $\bar{F}_{\nu}=-\partial H_{e}(\vec{R};\vec{r}_{T},\vec{p}_{T})/\partial R_{\nu}$.
The integral, then, gives the average force produced by a \emph{set}
of trajectories with each one contributing to the force with a weight
$\mathrm{d}\vec{r}\,\mathrm{d}\vec{p}\,\rho_{e}(\vec{r},\vec{p})$.

This averaging has the effect of reducing the ability of electrons
to pass energy to the nuclei. To understand this, consider the following
simple model. Suppose we have one nucleus that experiences forces
from the electrons that have two contributions: $\vec{F}=-k\vec{X}+\vec{f}(t)$.
The first, harmonic, term corresponds to motion on a Born-Oppenheimer
surface. The small residual non-adiabatic corrections are represented
by the force $\vec{f}(t)$. Let us define the energy of the nucleus
as $U_{N}=P^{2}/2M+kX^{2}/2$. The rate change of this energy (the
power given to the nucleus) satisfies $\dot{U}_{N}=\dot{\vec{X}}\cdot\vec{f}(t)$.
If we now evaluate the average power supplied over some time interval
$\tau$ we obtain\begin{equation}
\left\langle \dot{U}_{N}\right\rangle _{\tau}=\frac{1}{\tau}\int_{t_{0}}^{t_{0}+\tau}\,\dot{\vec{X}}\cdot\vec{f}(t)\,\mathrm{d}t\label{eq:efest05}\end{equation}
If $\vec{f}(t)$ varies only slightly during a vibrational period
of the nucleus, then the power will be very small, as the average
velocity of the nucleus is zero. If it fluctuates rapidly but there
are no correlations between $\dot{\vec{X}}$ and $\vec{f}$, then
the fluctuations average out and the power again vanishes. However
if we allow $\dot{\vec{X}}$ to be a functional of $\vec{f}$, via
the equation of motion, then in Eq. \ref{eq:efest05} there are correlated
force-velocity fluctuations, which can in turn be expressed in terms
of the force-force correlation function\cite{horsfield-2004-a}. It
is these correlated fluctuations that generate the power. This result
is known as the fluctuation-dissipation theorem. The effect of the
averaging over electronic trajectories in the Ehrenfest approximation
is to break the microscopic correlations between the force experienced
by the nucleus due to the electrons and the momentum of the nucleus,
and thus to suppress the energy that can be transferred to the nucleus.
This breaking of correlation is caused by the averaging of the electronic
trajectories for every nuclear configuration.

The above analysis is most natural for metallic systems which we can
think of as nuclei embedded in a gas of rapidly moving light electrons.
For small molecules with well spaced energy levels the criticism of
the Ehrenfest approximation can be made differently. The calculations
usually involve an electronic transition between two levels, induced
by the momentum of the nuclei. It is found that the results are often
in poor agreement with accurate quantum calculations. This is explained
by noting that the single nuclear trajectory experiences forces simultaneously
from a number of partially occupied energy surfaces, whereas it should
be experiencing it from one only, but with separate trajectories on
each surface. This is just another manifestation of the averaging
discussed above.

These problems manifest themselves in the failure to produce certain
outcomes in simulations, such as heating of nuclei by current carrying
electrons\cite{horsfield-2004-a}, thermal equlibrium between electrons
and nuclei\cite{theilhaber-1992-a,parandekar-2005-a}, and correct
transition probablilities\cite{hack-2000-a}. Various schemes have
been proposed to overcome these problems. One starting point is to
recognise that the ionic wavefunction rapidly decoheres after a transition
leading to modified equations of motion\cite{hack-2001-a,zhu-2004-a,zhu-2004-b},
or to a stochastic algorithm in which trajectories on separate Born-Oppenheimer
energy surfaces are treated as independent (see Section \ref{sub:Surface-hopping-methods}
below). Another is to treat the electrons and nuclei as having correlated
motion leading to wavepacket-like methods\cite{diestler-1983-a,kilin-2004-a}
(see Section \ref{sub:Correlated-electron-ion-dynamics} below).

While the Ehrenfest approximation is often presented in the literature
as a straw man (that is, a method used simply to show how much better
other methods are), it is also used to obtain useful results. Two
applications are:

\begin{enumerate}
\item Time-dependent density functional theory has been used successfully
to study the friction forces experienced by hydrogen atoms scattering
off a metallic surface \cite{baer-2004-a}. The friction is due to
the excitation of electrons by the moving atoms. Energy transfer in
this direction is correctly described by the Ehrenfest approximation.
\item Polymers have been successfully studied quite extensively using the
Ehrenfest approximation within a simple tight binding model for the
electrons. Investigations have included: polaron drift in an applied
electric field \cite{johansson-2004-a}; the dynamics of polarons
formed after the excitation of electrons by photons \cite{an-2004-a};
the nucleation of stable self-localised excitations \cite{block-1996-a}.
\end{enumerate}

\subsubsection{\label{sub:Surface-hopping-methods}Surface hopping methods}

Surface hopping was originally proposed by Tully and Preston \cite{tully71}
as a phenomenological extension of the classical trajectory method
which incorporates non-adiabatic effects. In the classical trajectory
method, an ensemble of trajectories sampling a series of initial conditions
for a gas phase chemical reaction is integrated in time. Quantities
such as reaction cross sections are obtained as averages over the
ensemble.

As mentioned in the previous section, within the Ehrenfest approximation
nuclei feel a force from the electrons that is an average over many
electronic trajectories. In cases where adiabatic electronic states
are close to each other in energy over relatively small regions of
space and are well separated otherwise, as in the case of a non-adiabatic
chemical reaction, when the nuclei leave a region of strong coupling
the force felt by them is a weighted average of the force in each
surface. Tully and Preston \emph{}observe that \emph{\char`\"{}intuition
requires that the motion take place on one surface or the other\char`\"{}}.
That is, after going through a region where there is significant coupling
between surfaces the ions must stay on the original surface on which
they were moving or jump to a different one. This intuition is guided
by the fact that in a molecular beam experiment, reactants convert
to one set of products or the other, but one does not detect states
which are mixture of different product channels. This implies some
decoherence of any mixed state before the outcome is measured (or
during the measurement), which in turn suggests a picture of the dynamics
in which nuclei move adiabatically unless they encounter a region
of strong non-adiabatic coupling. This can be either a point where
adiabatic surfaces cross or where they are close in energy.

This idea was implemented into the first realization of the surface
hopping procedure by determining a fixed region or set of points in
configuration space where switching between adiabatic states is likely.
When the system passes through those regions during the dynamics the
decision to hop to a different surface is taken according to a probability
calculated from the Landau-Zener expression (see section \ref{sub:Landau-Zener-theory}).
After a jump to a different potential energy surface has taken place,
the nuclei change their potential energy by a finite amount, which
is small if the jump occurs when the two surfaces are close to each
other. In order to conserve energy the surface hopping procedure must
be complemented with a prescription on how to change the velocities
of the nuclei after a hop. In their original proposal, Tully and Preston
rescaled the velocities in one particular direction, chosen from the
topology of the potential energy surface and the calculated couplings.
As with the standard classical trajectory method, properties of interest
are calculated from the properties of an ensemble of integrated trajectories.

Tully later extended the method by lifting the constraint of fixed
hopping regions and named the new method Molecular Dynamics with Quantum
Transitions (MDQT) \cite{tully90,hammesschiffer94}. In this method,
the jumps between surfaces can occur at any point during the trajectory.
The decision whether a hop should occur and to which surface is taken
according to a probability of a jump occurring. This is calculated
from the time evolving state populations, which in turn is obtained
from the simultaneous integration of the nuclear dynamics on a certain
adiabatic potential energy surface and the equation of motion for
the electronic wave function or density matrix. This starts by recasting
the equations for Ehrenfest dynamics given before in terms of the
adiabatic electronic states (see section \ref{sec:The-electron-phonon-Hamiltonian}).
First, the following anzatz is made for the electronic wavefunction
\[
\Psi(t)=\sum_{j}c_{j}(t)\Phi_{j}(\vec{R}(t))\]
 where $\Phi_{n}(\vec{R}(t))$ are the adiabatic eigenstates of the
instantaneous electronic Hamiltonian, and are a function of time via
the nuclear coordinates. Replacing this ansatz in the Schr\"{o}dinger
equation for the electrons we obtain the following equation of motion
for the electronic wavefunction coefficients: \begin{equation}
i\hbar\dot{c}_{j}=c_{j}E_{j}(\vec{R}(t))-i\hbar\sum_{k}c_{k}\dot{R}\cdot\vec{d}_{kj}\label{sheom1}\end{equation}
 where the \emph{non-adiabatic coupling vector} $\vec{d}_{kj}$ is
defined by \[
\vec{d}_{kj}=\int\Phi_{k}^{*}(\vec{R}(t))\nabla_{\vec{R}}\Phi_{j}(\vec{R}(t))\, d\vec{R}.\]
 The same result can be obtained via a variational procedure from
a suitably constructed Lagrangian \cite{todorov-2001-a}. The second
term in Eq. \ref{sheom1} is the one responsible for transitions between
states. This transitions will be more effective if the nuclear velocity
points in the direction of the non-adiabatic coupling vector. The
rate of change of the population on a given state is given by the
equation of motion for a diagonal element of the electronic density
matrix: \begin{equation}
\dot{\rho}_{ii}=-i\hbar\sum_{j}2{\rm Re}\left\{ \rho_{ij}\vec{R}\cdot\vec{d}_{ij}\right\} =\sum_{j}b_{ij}\label{masteq}\end{equation}
 Where we have taken into account the fact that the matrix of non-adiabatic
coupling vectors is anti-hermitian. The fewest switches algorithm
proposed by Tully was designed so that the choice of where to hop
resembles the kinetic Monte Carlo integration of a master equation
for the state occupations given by Eq. \ref{masteq}. In a given time
step the probability that a trajectory occupying state $i$ hops to
a different state $j$ is proportional to $b_{ij}$. This probability
is constrained in a way that minimizes the number of hops required
to achieve consistency between the occupations and the number of trajectories
in the ensemble that occupy a given state. Tully demonstrates that
the algorithm ensures that at any time the number of trajectories
in a given state is representative of the occupation of that state.
However, the problem with this argument is that there is no unique
set of occupations, but instead there is one for each trajectory.
If the occupations of the states from different trajectories diverge
fast relative to one another because the nuclei follow different paths
(something that will occur faster the more hops there are) then consistency
will be lost \cite{fang99}.

In order to conserve energy whenever a hop occurs the velocity component
pointing in the direction of the non-adiabatic coupling vector is
rescaled. If, from the algorithm, it is determined that a hop must
occur to a potential energy surface for which there is not enough
energy, the hop is aborted and the velocity in the direction of the
non-adiabatic coupling vector is reversed \cite{hammesschiffer94}.
These classically forbiden hops can become a problem if they occur
too often since they contribute to breaking the consistency between
the propagated occupations and the number of trajectories in a given
state \cite{fang99}.

Comparison with fully quantum results for simple systems shows that
the method produces qualitatively correct results \cite{topaler-1997-a,tully90}.
Unlike the Ehrenfest method it reproduces the correct Boltzman populations
when the system is coupled to a thermostat \cite{parandekar-2005-a}.
The fact that for each particular trajectory a full quantum wavefunction
is propagated means that the method accounts in some way for coherence
within each trajectory and some effects due to interference can be
reproduced \cite{tully90}. However no interference exists between
trajectories. Further, it is difficult to asses, because of the way
in which the interaction between nuclear and electronic degrees of
freedom is treated, whether the resulting quantum coherence effects
are meaningful \cite{topaler98}. One particular problem is the fact
that the method seems to work only when the electronic states are
represented in an adiabatic basis \cite{tully98}. This is probably
due to the fact that the use of this basis leads to the least number
of hops \cite{bach01}. The rescaling of the velocity after a hop,
which is another controversial feature of surface hopping, can be
justified in a number of ways by semiclassical arguments \cite{coker95,herman95}.
Although the replacement of the propagation of a single, fully coherent,
wavefunction by an ensemble of trajectories can be made plausible
by semiclassical arguments \cite{tully98} some aspects of the method
are clearly \emph{ad hoc} making it difficult to improve in a systematic
way, or to predict in which situations it can be applied successfully.

Extensions of the method exist for cases in which there are both discrete
and continuum states \cite{sholl98}. They either use different schemes
to choose when to hop \cite{blais83} or replace sudden hops by a
continuous switching \cite{volobuev00}. Prezhdo and Rossky proposed
a combination of the Ehrenfest and surface hopping procedures in which
the force used to propagate the nuclear trajectory comes from the
integration of Ehrenfest equations, not just one adiabatic state,
but this wavefunction is reduced to a single adiabatic state when
a hop occurs or when the validity of the Ehrenfest approximation is
violated according to some pre-established criteria \cite{prezhdo97}.

Tully's MDQT is one of the most used methods to deal with problems
in which coupling between electrons and ions is fundamental. In particular,
it has been extensively used to study non-adiabatic processes in liquids,
given that the complexity of liquid structure prevents application
of other techniques. One of the first applications of the method was
to the problem of the solvated electron. Space and Coker explored
the relaxation of an excess electron in dense fluid helium. MDQT provided
information on the dependence of the electronic relaxation process
on the initial electronic state \cite{space91}. Later, MDQT and some
of its variants were applied to an excess electron in water \cite{prezhdo96,drukker98,wong01}.
Another classical problem to which the method has been applied is
the simplest photochemical reaction in solution, photodissociation
of a diatomic molecule. Coker's group pioneered the application of
MDQT to this problem, studying the photodissociation of I$_{2}$ in
liquid and solid rare gases \cite{batista96,batista97}. Others later
studied photodissociation of diatomic molecules embedded in rare gas
clusters \cite{niv99,niv00}. The case of the ICN molecule which can
isomerize after dissociation has been studied in rare gas matrices
\cite{alberti00}, where some of the vibrational modes of the molecule
were included within the quantum description, in bulk water \cite{winter03}
and at the liquid/vapour interface of water \cite{winter04}. An important
effect of the solvent in the case of photodissociation is that some
processes which occur with high quantum yields in the gas phase are
inhibited in the solvent. This effect is explained in terms of the
caging of the reactants which are held together in the solvent and
have the opportunity to recombine. Particularly dramatic is the case
of azomethane (${\rm H_{3}CNNCH}_{3}$) which photodissociates in
the gas phase but isomerizes around the NN double bond in the solvent.
This problem was studied using surface hopping in the group of Persico
\cite{cattaneo98,cattaneo99,cattaneo01}, becoming one of the first
non-adiabatic simulations of condensed phase photochemistry with a
realistic solvent and a polyatomic solute. Since photodissociation
is a classic problem in photochemistry, there are abundant experimental
results for these systems which can be compared with the results obtained
from the simulations. In general the agreement has been found to be
good.

Relaxation of a charge transfer excited state via intramolecular electron
transfer in solution is another classic example of a non-adiabatic
problem in solution chemistry. Lobaugh and Rossky studied the relaxation
of excited betaine in acetonitrile. The electronic structure of the
molecule was described using configuration interaction applied to
a simple semiempirical model \cite{lobaugh99}. The authors found
that besides the motion of the solvent, which is effective during
the initial stages of the relaxation from the excited state, also
some internal motions of the molecule played a fundamental role in
the relaxation process.

Photoisomerization around a double bond is a paradigmatic example
of non-adiabatic chemistry, and important biological functions such
as vision hinge on it. Surface hopping has been used to study the
photoisomerization of butadiene \cite{ito97}, the chromophore of
the photoactive yellow protein \cite{groenhof04} and of retinal in
bacteriorhodopsin \cite{warshel01}. Azobenzenes are one class of
compounds that can be used for the generation of materials which change
their properties via illumination. \emph{Trans} to \emph{cis} isomerization
of azobenzene is the key for this effect, and surface hopping has
recently been used to solve some controversy on the detailed mechanism
of the process \cite{ciminelli04}.

Very few applications of surface hopping to processes in solids exist.
Yokozawa and Miyamoto studied the breaking of Si-H bonds from H terminated
O vacancies in ${\rm SiO}_{2}$ by hot electrons using a primitive
combination of the surface hopping procedure and first principles
molecular dynamics \cite{yokozawa00}. Bruening and Friedman used
surface hopping and a tight binding Hamiltonian to describe charge
transfer from a conducting polymer to C$_{60}$ molecules \cite{bruening97}.
Bach and Gross treated with surface hopping the problem of charge
transfer in molecule-surface scattering \cite{bach01}.

An important difficulty with the implementation of surface hopping
procedures is the need for accurate adiabatic states and non-adiabatic
coupling vectors which must be obtained from some model description
of the electronic structure or as a parametrization of \emph{ab initio}
results prior to the simulation. Recently, surface hopping has been
implemented in \emph{ab initio} molecular dynamics methods with on
the fly calculation of the electronic structure. Doltsinis and Marx
have implemented surface hopping within a restricted open-shell Kohn-Sham
scheme \cite{frank98} which allows the calculation of the low lying
excited states and the corresponding coupling vectors between states
\cite{doltsinis02}. The method was applied to \emph{cis-trans} photoisomerization
of formaldimine (${\rm H}_{2}{\rm CNH}$) \cite{doltsinis02}, excited
state proton transfer and internal conversion of \emph{o}-hydroxybenzaldehyde
\cite{doltsinis04}, and the photostability of methylated DNA bases
\cite{langer04}. Prezhdo \emph{et al.} have implemented a similar
scheme where the adiabatic states used for the surface hopping propagation
are different excited Kohn-Sham determinants and applied the method
to study the nonradiative relaxation of the chromophore of the green
fluorescent protein and electron injection from alizarin into titanium
dioxide \cite{craig05}.

Recently, some new formulations of surface hopping techniques have
appeared which are constructed from well defined approximations of
the exact quantum dynamics \cite{herman05,thorndyke05,nielsen-2000-a}.
These techniques have up to now been demonstrated to produce excellent
results when compared with exact calculations for model systems, and
some small problems \cite{thorndyke05}. Perhaps in the future these
formally correct methods will become available for realistic condensed
phase systems.

\subsubsection{\label{sub:frozen-gaussians}Frozen Gaussian Approximation and related
methods}

In this section, we will cover the Frozen Gaussian approximation (FGA)
and various related semiclassical (SC) techniques including the Initial
Value Representation (IVR) and the Herman-Kluk propagator. We will
also consider methods which use the FGA for nuclear wavefunctions
(one of these, surface hopping, which sometimes uses FGA for the nuclei
is discussed in Section \ref{sub:Surface-hopping-methods}). There
are various other reviews of these techniques which go into more detail:
a thorough, though very formal, review of semiclassical methods\cite{Baranger2001a};
an overview of SC-IVR methods\cite{Miller2001a}; a discussion of
wavepacket (FGA) and surface hopping methods\cite{Herman1994a}; and
a more general review covering time dependent methods for large systems\cite{makri99}.

The use of gaussian wavepackets as a basis for nuclear wavefunctions
started in scattering calculations which relegated the electrons to
the role of providing a potential; their use was pioneered by Heller\cite{heller-1975-a,heller-1976-a,Heller1976b,heller-1977-a,heller-1977-b,davis-1979-a}.
From this work, he proposed that a simple way to broaden single classical
trajectories would be to use a Gaussian function whose width was fixed
and whose average position and momentum followed that trajectory\cite{heller-1981-a}.
For a gaussian centred on the phase-space point $(\mathbf{r},\mathbf{p})$
with width $\gamma$, we write:

\begin{equation}
\langle\mathbf{x}|\mathbf{r},\mathbf{p}\rangle=\left(\frac{2\gamma}{\pi}\right)^{N/4}\exp\left[-\gamma(\mathbf{x}-\mathbf{r})^{2}+{\rm i}\mathbf{p}\cdot(\mathbf{x}-\mathbf{r})/\hbar\right]\label{eq:FG02}\end{equation}

The semiclassical approximation (due originally to Van Vleck\cite{VanVleck1928a}
as an extension of the WKB method to time-dependent problems) provides
a way to address dynamics. Consider the amplitude to change from state
1 to state 2 for a system with continuous position and momentum $(\mathbf{r},\mathbf{p})$:\begin{equation}
T_{1\rightarrow2}=\langle\Psi_{2}\left|e^{-{\rm i}\hat{H}t/\hbar}\right|\Psi_{1}\rangle=\int\int{\rm d}\mathbf{r}_{1}{\rm d}\mathbf{r}_{2}\Psi_{2}^{\star}(\mathbf{r}_{2})\left\langle \mathbf{r}_{2}\right|e^{-{\rm i}\hat{H}t/\hbar}\left|\mathbf{r}_{1}\right\rangle \Psi_{1}(\mathbf{r}_{1})\label{eq:FG04}\end{equation}
 The semiclassical approximation changes the propagator to: \begin{eqnarray}
\langle\mathbf{r}_{2}\left|e^{-{\rm i}\hat{H}t/\hbar}\right|\mathbf{r}_{1}\rangle & = & \sum_{\mathrm{roots}}Ce^{{\rm i}S(\mathbf{r}_{2},\mathbf{r}_{1})/\hbar}\label{eq:FG05}\\
S(\mathbf{r}_{2},\mathbf{r}_{1}) & = & \int_{0}^{t}dt^{\prime}\left(T(t^{\prime})-V(t^{\prime})\right)\nonumber \end{eqnarray}
 Here $S(\mathbf{r}_{2},\mathbf{r}_{1})$ is the classical action
for the trajectory going from $\mathbf{r}_{1}$ to $\mathbf{r}_{2}$
in time $t$; this makes clear the connection to Feynman's path-integral
formulation of quantum mechanics. The factor $C$ contains a number
of factors which there is not space to discuss here (see for instance
Ref.\cite{Sun1997a,Miller2001a}). The sum over roots arises from
the stationary phase limit\cite{Pechukas1969a} which gives all classical
paths for which $\delta S=0$. Solving this equation as written gives
a non-linear boundary value problem, where all values of $\mathbf{p}_{1}$
which yield $\mathbf{r}_{2}$ must be found; this will lead, in general,
to multiple roots.

The initial value representation replaces the integration over final
coordinates, $d\mathbf{r}_{2}$, with integration over initial momenta,
$d\mathbf{p}_{1}$ (bringing in a Jacobian with the change of variables).
The sum over roots also disappears because the initial conditions
determine a unique classical trajectoty. We can write:

\begin{equation}
\sum_{\mathrm{roots}}\int{\rm d}\mathbf{r}_{2}=\int{\rm d}\mathbf{p}_{1}\left|\frac{\partial\mathbf{r}_{2}}{\partial\mathbf{p}_{1}}\right|\label{eq:FG06}\end{equation}
 This approximation allows some quantum interference and tunnelling
effects to be described, while involving only real, classical trajectories.

The most commonly used combination of a Frozen Gaussian basis, and
the semiclassical IVR is the Herman-Kluk propagator\cite{Herman1984a,kluk-1985-a}.
\begin{eqnarray*}
\psi(\mathbf{x},t) & = & \int\frac{{\rm d}^{N}\mathbf{r}_{i}{\rm d}^{N}\mathbf{p}_{i}}{(2\pi\hbar)^{N}}\langle\mathbf{x}|\mathbf{r}_{t},\mathbf{p}_{t}\rangle C(\mathbf{r}_{i},\mathbf{p}_{i},t)\\
 & \times & \exp\left[iS(\mathbf{r}_{i},\mathbf{p}_{i},t)/\hbar\right]\langle\mathbf{r}_{i},\mathbf{p}_{i}|\psi_{0}\rangle,\end{eqnarray*}
 for an $N$-dimensional problem. A trajectory starts at phase space
point $(\mathbf{r}_{i},\mathbf{p}_{i})$ and runs for time $t$ to
a phase space point $(\mathbf{r}_{t},\mathbf{p}_{t})$. The form of
the semiclassical prefactor required is important if the calculation
is to be carried over long time periods\cite{kluk-1985-a}: \begin{equation}
C(\mathbf{r}_{i},\mathbf{p}_{i},t)=\left|\frac{1}{2}\left(\frac{\partial\mathbf{p}_{t}}{\partial\mathbf{p}_{i}}+\frac{\partial\mathbf{r}_{t}}{\partial\mathbf{r}_{i}}-2\gamma i\hbar\frac{\partial\mathbf{r}_{t}}{\partial\mathbf{p}_{i}}+\frac{i}{2\gamma\hbar}\frac{\partial\mathbf{p}_{t}}{\partial\mathbf{r}_{i}}\right)\right|^{1/2}\label{eq:FG03}\end{equation}

The SC wavepacket methods described thus far have the important features
of time-reversal and unitarity\cite{Herman1986a}. However, in their
original form, there are often rapid oscillations in the integrand
which cause problems with convergence. These oscillations were shown
to be significantly damped\cite{Walton1996a} by the merging of the
cellular dynamics method\cite{Heller1991a} with the Herman-Kluk propagator.
Other solutions to this problem include time integration over short
periods\cite{Elran1999a} yielding effective averaging over the short
period. Still within the realm of reactive scattering, the limitations
of the Herman-Kluk propagator have been explored in 2D and 4D modelling
of H$_{2}$ scattering from Cu(001) \cite{McCormack1999a}. The 2D
simulation was not sufficiently accurate, while the full 4D simulation
required inordinate amounts of computational effort. There has been
application to non-adiabatic electronic evolution, in particular to
the canonical problem of electron solvation\cite{Schwartz1996a}.
In this area, FG propagation coupled with perturbation theory has
been compared to the surface-hopping techniques\cite{Neria1993a}.
In regimes where tunnelling is important, surface hopping is less
accurate, though the FG propagation is only accurate over short timescales,
and can depend on the width chosen for the Gaussian. Some analysis
of the accuracy of using classical mechanics for the propagation of
the wavepackets in condensed many-body systems\cite{Herman1999a}
showed that while the wavefunction can spread (losing the justification
for the FGA), densities continue to be localised, due to interference
effects. These ideas have been applied to the calculation of non-linear
optical response functions, with reasonable accuracy\cite{Noid2003a}.
These SC-IVR methods scale exponentially with the number of degrees
of freedom\cite{Brewer1999a} though this does depend on the number
of final states contributing and the energy resolution required.

In order to model non-adiabatic transitions, the FGA (and any SC-IVR)
must be extended. It has been shown\cite{Sun1997a,Sun1998a} that
it is possible to linearise the SC-IVR\cite{Miller1970b}, splitting
the entire system into a system and a bath; and expansion is then
made in terms of the bath coordinates, using Wigner distributions.
This technique has been applied to simple models (a Morse oscillator
coupled to a single harmonic mode) \cite{Sun1997a}, and the spin
boson problem with different spectral weights\cite{Sun1998a,Wang1999a},
with good agreement to exact results.

The multiple-spawning technique\cite{Martinez1996a,Martinez1997a,BenNun1998a}
uses frozen Gaussians as basis functions during a time-dependent simulation.
An effective non-adiabatic coupling is monitored between two states
$I$ and $I^{\prime}$; using a diabatic (see \ref{sec:The-electron-phonon-Hamiltonian})
representation for the electronic wavefunctions, this can be written
as :

\begin{equation}
d(\mathbf{R})=\left|\frac{\langle I|\hat{H}|I^{\prime}\rangle}{V_{II}(\mathbf{R})-V_{I^{\prime}I^{\prime}}(\mathbf{R})}\right|\label{eq:FG07}\end{equation}
 When this coupling for state $I$ exceeds a threshold, new nuclear
wavefunctions are added (or \emph{spawned}) on the new state $I^{\prime}$
at a constant rate (so that more wavefunctions are added the longer
the system remains in the area of non-adiabatic coupling). These new
functions are then propagated, and can themselves spawn. This leads
to a system whose nuclear degrees of freedom follow multiple trajectories
on different surfaces simultaneously. It has been applied successfully,
among other things, to a two dimensional non-reactive collision\cite{BenNun1998a},
and more recently to light-driven reactions\cite{Martinez2005a}.

The Zhu-Nakamura (ZN) theory\cite{zhu-1994-a} for non-adiabatic transitions
(an extension of the Landau-Zener formalism that works with adiabatic
states) has been combined with the use of frozen Gaussians for the
nuclear wavefunctions\cite{Zhu2001b,Zhu2001a,Kondorskiy2004a} to
extend both theories, both with the surface hopping formalism\cite{Zhu2001b,Zhu2001a}
(specifically using ZN theory to calculate the non-adiabatic transitions)
and as a separate method\cite{Kondorskiy2004a}. This approach has
the advantage of giving the rates from an analytic theory, while allowing
nuclear wavefunction propagation, and works well in multidimensional
problems (by comparison to more complicated and costly numerical calculations
of the problems).

\subsubsection{\label{sub:Correlated-electron-ion-dynamics}Correlated electron-ion
dynamics}

\begin{figure}
\begin{center}\includegraphics[%
  width=0.6\columnwidth,
  keepaspectratio]{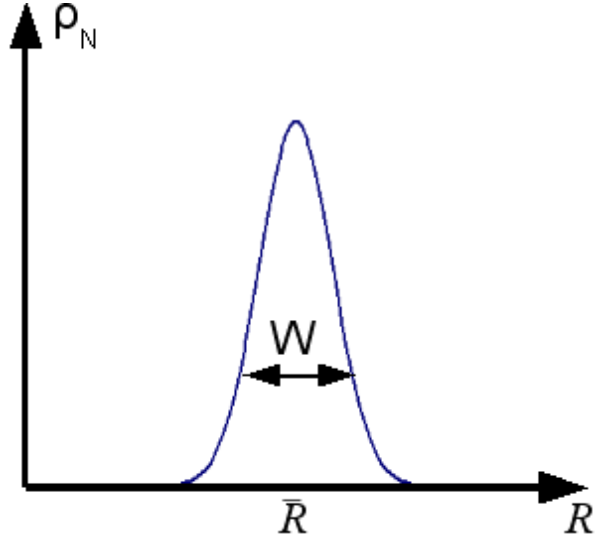}\end{center}

\caption{\label{cap:The-central-idea}The central idea in correlated electron
ion dynamics is that the width of a nuclear wave packet is narrow
on the length scale of the separation between nuclei. That is, $W\ll d$
where $d$ is a typical distance between nuclei. In this figure the
mean density of the nuclei as a function of position $R$ is given
the symbol $\rho_{N}$. Correlations between the electrons and nuclei
are maintained by allowing the electronic state to vary as a function
of position of the nuclei within the wave packet.}
\end{figure}

The small amplitude moment expansion is a scheme to introduce the
correlations between the electronic fluctuations and the nuclei that
are missing in the Ehrenfest approximation\cite{horsfield-2004-a}.
When this expansion is combined with molecular dynamics, we refer
to it as Correlated Electron-Ion Dynamics (CEID). The starting point
is the Ehrenfest equations (which are, of course exact, and which
give the name to the approximation):\begin{eqnarray}
\dot{\bar{R}}_{\nu} & = & \frac{\bar{P}_{\nu}}{M_{\nu}}\nonumber \\
\dot{\bar{P}}_{\nu} & = & \bar{F}_{\nu}\nonumber \\
\bar{F}_{\nu} & = & -\mathrm{Tr}\left\{ \hat{\rho}\frac{\partial\hat{H}_{e}}{\partial\hat{R}_{\nu}}\right\} \label{eq:same00}\end{eqnarray}
where $\bar{R}_{\nu}=\mathrm{Tr}\left\{ \hat{R}_{\nu}\hat{\rho}\right\} $
and $\bar{P}_{\nu}=\mathrm{Tr}\left\{ \hat{P}_{\nu}\hat{\rho}\right\} $,
and $\hat{R}_{\nu}$ and $\hat{P}_{\nu}$ are the position and momentum
operators for the nuclei. In the Ehrenfest approximation we were able
to work with a single trajectory: in that case $\bar{R}_{\nu}=R_{T,\nu}$
and $\bar{P}_{\nu}=P_{T,\nu}$. However, since the Ehrenfest equations
refers to quantum nuclei, the uncertainty principle makes it impossible
to have individual trajectories. The best we can manage are narrow
wave packets (see figure \ref{cap:The-central-idea}). In this case
$\bar{R}_{\nu}$ and $\bar{P}_{\nu}$ correspond to the trajectory
of the mean of the wave packet.%

If the wave packets really are narrow (which they often are in condensed
systems), then we can expand $\hat{H}_{e}(\hat{R})$ about $\bar{R}$
in the following way\cite{horsfield-2004-b} (where $\hat{R}$ and
$\bar{R}$ stand for the set of coordinates $\left\{ \hat{R}_{\nu}\right\} $
and $\left\{ \bar{R}_{\nu}\right\} $ respectively):\begin{equation}
\fl\hat{H}_{e}(\hat{R})=\hat{H}_{e}(\bar{R})+\sum_{\nu}\Delta\hat{R}_{\nu}\frac{\partial\hat{H}_{e}(\bar{R})}{\partial\bar{R}_{\nu}}+\frac{1}{2}\sum_{\nu\nu'}\Delta\hat{R}_{\nu}\Delta\hat{R}_{\nu'}\frac{\partial^{2}\hat{H}_{e}(\bar{R})}{\partial\bar{R}_{\nu}\partial\bar{R}_{\nu'}}+\cdots\label{eq:same01}\end{equation}
where $\Delta\hat{R}_{\nu}=\hat{R}_{\nu}-\bar{R}_{\nu}$. Then the
mean force (Eq. \ref{eq:same00}) satisfies\begin{equation}
\fl\bar{F}_{\nu}=-\mathrm{Tr}_{e}\left\{ \hat{\rho}_{e}\frac{\partial\hat{H}_{e}(\bar{R})}{\partial\bar{R}_{\nu}}\right\} -\sum_{\nu'}\mathrm{Tr}_{e}\left\{ \hat{\mu}_{1,\nu'}\frac{\partial^{2}\hat{H}_{e}(\bar{R})}{\partial\bar{R}_{\nu}\partial\bar{R}_{\nu'}}\right\} -\frac{1}{2}\sum_{\nu'\nu''}\mathrm{Tr}_{e}\left\{ \hat{\mu}_{2,\nu'\nu''}\frac{\partial^{3}\hat{H}_{e}(\bar{R})}{\partial\bar{R}_{\nu}\partial\bar{R}_{\nu'}\partial\bar{R}_{\nu''}}\right\} +\cdots\label{eq:same02}\end{equation}
where $\hat{\rho}_{e}=\mathrm{Tr}_{N}\left\{ \hat{\rho}\right\} $,
$\hat{\mu}_{1,\nu}=\mathrm{Tr}_{N}\left\{ \hat{\rho}\Delta\hat{R}_{\nu}\right\} $,
and $\hat{\mu}_{2,\nu\nu'}=\mathrm{Tr}_{N}\left\{ \hat{\rho}\Delta\hat{R}_{\nu}\Delta\hat{R}_{\nu'}\right\} $.
The first term in Eq. \ref{eq:same02} is the Ehrenfest approximation.
The higher terms account for the fact that the wave packet has some
finite width, and that the mean force must include an average over
the paths included within the packet. However, at this point it is
important to note a fundamental distinction. There are \emph{two}
separate contributions to the total width of the wave packet. The
more obvious is just the normal quantum width of the nucleus, and
this contributes to $\hat{\mu}_{2,\nu\nu'}$ and higher moments. The
other contribution is due to transitions between Born-Oppenheimer
surfaces. If a nucleus starts a trajectory on one particular Born-Oppenheimer
energy surface, and there is some coupling to another surface, then
that trajectory will split into two trajectories, one on each surface.
Since the forces experienced on each surface can be different the
two trajectories can deviate from one another, producing broadening
of the total wave packet. The first moment ($\hat{\mu}_{1,\nu}$)
contains only this contribution, but it appears in all other moments
as well.

We now seek to understand the meaning of the moments. First we note
that $\hat{\rho}_{e}$, $\hat{\mu}_{1,\nu}$ and $\hat{\mu}_{2,\nu\nu'}$
are electronic operators (the traces have only been taken over the
nuclear degrees of freedom). We can therefore take matrix elements
with respect to electronic states. Suppose we have energy surfaces
$\alpha$ characterised by a set of electronic states $|\alpha\rangle$.
We then have\begin{eqnarray}
\rho_{e,\alpha\alpha} & = & \langle\alpha|\hat{\rho}_{e}|\alpha\rangle=\int\rho_{\alpha\alpha}(\vec{R})\mathrm{d}\vec{R}\nonumber \\
\mu_{1,\nu,\alpha\alpha} & = & \langle\alpha|\hat{\mu}_{1,\nu}|\alpha\rangle=\int\rho_{\alpha\alpha}(\vec{R})\,(R_{\nu}-\bar{R}_{\nu})\,\mathrm{d}\vec{R}\nonumber \\
\mu_{2,\nu\nu',\alpha\alpha} & = & \langle\alpha|\hat{\mu}_{2,\nu\nu'}|\alpha\rangle=\int\rho_{\alpha\alpha}(\vec{R})\,(R_{\nu}-\bar{R}_{\nu})(R_{\nu'}-\bar{R}_{\nu'})\,\mathrm{d}\vec{R}\label{eq:same03}\end{eqnarray}
where $\rho_{\alpha\alpha}(\vec{R})=\langle\alpha\vec{R}|\hat{\rho}|\alpha\vec{R}\rangle$,
and is the ionic density projected onto surface $\alpha$. Thus $\rho_{e,\alpha\alpha}$
is just the probability of being on surface $\alpha.$ If we define
the mean value of some observable $\hat{Q}$ on surface $\alpha$
to be $\langle\alpha|\hat{Q}|\alpha\rangle/\rho_{e,\alpha\alpha}$
then $\mu_{1,\nu,\alpha\alpha}$ just equals the mean position of
the nucleus on the surface, measured relative to the mean trajectory
$\bar{R}_{\nu}$, multiplied by the probability of being on that surface.
Similarly, $\mu_{2,\nu\nu',\alpha\alpha}$ gives the width of the
packet moving on surface $\alpha$, multiplied by the probability
of being on that surface, and thus is a measure of the quantum width
of the nucleus.

Thus, in summary, the small amplitude moment expansion corrects the
Ehrenfest approximation by allowing different trajectories on different
energy surfaces, and by giving the nuclei a finite quantum width.
To make this into a practical method we need some efficient way of
evaluating the moments $\hat{\mu}_{1,\nu}$, $\hat{\mu}_{2,\nu\nu'}$,
\emph{etc}. This is achieved by integrating their equations of motion\cite{horsfield-2004-b}.
These equations follow from the Liouville equation (Eq. \ref{eq:efest-01}),
and the definition of the moments. Thus we have\begin{eqnarray}
\fl\frac{\mathrm{d}\hat{\rho}_{e}}{\mathrm{d}t} & = & \frac{1}{\mathrm{i}\hbar}\left[\hat{H}_{e}(\bar{R}),\hat{\rho}_{e}\right]-\frac{1}{\mathrm{i}\hbar}\sum_{\nu}\left[\hat{F}_{\nu},\hat{\mu}_{1,\nu}\right]+\cdots\nonumber \\
\fl\frac{\mathrm{d}\hat{\mu}_{1,\nu}}{\mathrm{d}t} & = & \frac{\hat{\lambda}_{1,\nu}}{M_{\nu}}+\frac{1}{\mathrm{i}\hbar}\left[\hat{H}_{e}(\bar{R}),\hat{\mu}_{1,\nu}\right]+\cdots\nonumber \\
\fl\frac{\mathrm{d}\hat{\lambda}_{1,\nu}}{\mathrm{d}t} & = & \frac{1}{2}\left(\Delta\hat{F}_{\nu}\hat{\rho}_{e}+\hat{\rho}_{e}\Delta\hat{F}_{\nu}\right)-\sum_{\nu'}\frac{1}{2}\left(\hat{K}_{\nu\nu'}\hat{\mu}_{1,\nu'}+\hat{\mu}_{1,\nu'}\hat{K}_{\nu\nu'}\right)+\frac{1}{\mathrm{i}\hbar}\left[\hat{H}_{e}(\bar{R}),\hat{\lambda}_{1,\nu}\right]+\cdots\label{eq:same04}\end{eqnarray}
where $\hat{F}_{\nu}=-\partial\hat{H}_{e}(\bar{R})/\partial\bar{R}_{\nu}$,
$\Delta\hat{F}_{\nu}=\hat{F}_{\nu}-\bar{F}_{\nu}$, $\hat{K}_{\nu\nu'}=\partial^{2}\hat{H}_{e}(\bar{R})/\partial\bar{R}_{\nu}\partial\bar{R}_{\nu'}$,
$\hat{\lambda}_{1,\nu}=\mathrm{Tr}_{N}\left\{ \hat{\rho}\Delta\hat{P}_{\nu}\right\} $
and $\Delta\hat{P}_{\nu}=\hat{P}_{\nu}-\bar{P}_{\nu}$. We can obtain
a rather straightforward interpretation of these equations by considering
matrix elements of the moments with respect to the electronic states
$|\alpha\rangle$, provided $\hat{H}_{e}(\bar{R})|\alpha\rangle={\rm i}\hbar\partial|\alpha\rangle/\partial t$.
Under these conditions from Eq. \ref{eq:same04} we obtain\begin{eqnarray}
\frac{\mathrm{d}\rho_{e,\alpha\alpha}}{\mathrm{d}t} & = & -\frac{1}{\mathrm{i}\hbar}\sum_{\nu\alpha'}\left(F_{\nu,\alpha\alpha'}\mu_{1,\nu,\alpha'\alpha}-\mu_{1,\nu,\alpha\alpha'}F_{\nu,\alpha'\alpha}\right)+\cdots\nonumber \\
\frac{\mathrm{d}\mu_{1,\nu,\alpha\alpha}}{\mathrm{d}t} & = & \frac{\lambda_{1,\nu,\alpha\alpha}}{M_{\nu}}+\cdots\nonumber \\
\frac{\mathrm{d}\lambda_{1,\nu,\alpha\alpha}}{\mathrm{d}t} & = & \Delta F_{\nu,\alpha\alpha}\rho_{e,\alpha\alpha}+\frac{1}{2}\sum_{\alpha'(\ne\alpha)}\left(\Delta F_{\nu,\alpha\alpha'}\rho_{e,\alpha'\alpha}+\rho_{e,\alpha\alpha'}\Delta F_{\nu,\alpha'\alpha}\right)+\cdots\label{eq:same05}\end{eqnarray}
The first equation (for the diagonal elements of the electronic density
matrix) tells us that transitions between Born-Oppenheimer surfaces
are driven by force couplings between the surfaces. If we think of
$\mu_{1,\nu,\alpha\alpha}$ as the position of a nucleus on the surface
multiplied by the probability of being on the surface, and $\lambda_{1,\nu,\alpha\alpha}$
as the momentum of a nucleus on the surface multiplied by the probability
of being on the surface (see Eq. \ref{eq:same03}), then the second
equation just gives the usual relation between velocity and momentum
(provided the probability of being on the surface is not changing).
Finally, the third equation relates the rate of change of momentum
of the nucleus on the surface to the force that it is experiencing
(the diagonal force term), plus some additional corrections derived
from the non-adiabatic interactions. Thus we see that the small amplitude
moment expansion at lower orders is describing classical trajectories
on multiple Born-Oppenheimer surfaces.

For a practical implementation of the scheme we need to truncate the
infinite hierarchy of equations of motion for the moments. If we work
with a fixed order of expansion for the Hamiltonian (in practice dropping
cubic and higher terms in the Taylor expansion in Eq. \ref{eq:same01}),
then for the highest order moments for which we have equations of
motion there will be additional moments appearing in those equations
for which there is no corresponding equation of motion. To produce
closure we therefore need to estimate these higher moments from the
lower ones already evaluated. For metallic systems, where nuclear
wave packets mostly breathe without splitting, satisfactory results
have been obtained\cite{todorov-2005-a,horsfield-2005-a} by a mean
field approximation which relates $\hat{\mu}_{2,\nu\nu'}$ and other
second moment operators to products of their own traces and $\hat{\rho}_{e}$.
This procedure is a simple case of a general formalism which is currently
under development. First we reconstruct an approximate density matrix
from the known moments, then we perform the necessary traces with
this density matrix to produce the required additional moments. This
process is made much easier because of two remarkable theorems. Consider
a density matrix for which a Wigner transform has been carried out
over the nuclear decrees of freedom (see \ref{sec:The-Wigner-transform}),
to give $\hat{\rho}_{W}(\vec{R}\vec{P})$ which is an electronic operator
which depends parametrically on the nuclear positions and momenta.
For one nuclear degree of freedom the first theorem states\cite{hillery-1984-a}
that\begin{equation}
\fl\int R^{n}P^{m}\hat{\rho}_{W}(RP)\,{\rm d}R\,{\rm d}P=\left(\frac{1}{2}\right)^{n}\sum_{l=0}^{n}C_{l}^{n}{\rm Tr}\left\{ \hat{\rho}\hat{R}^{l}\hat{P}^{m}\hat{R}^{n-l}\right\} \label{eq:XC01}\end{equation}
which relates the moments of the Wigner matrix to those of the density
matrix. The second theorem states that for quadratic Hamiltonians
the moments appearing in CEID can always be written in the form given
by the right-hand side of Eq. \ref{eq:XC01}. In short, CEID moments
are moments of the Wigner function. This has two immediate consequences.
First, we do not have to worry about the order of the operators in
the moments, so that we can completely characterise the moments by
the powers of position and momentum. Second, since the Wigner function
is a function of scalars we can approximate this function using standard
techniques, rather than attempting the much more complex process of
approximating functions of operators. For CEID we use harmonic oscillator
eigenfunctions to expand the Wigner function (an approach which has
been used in other similar contexts\cite{billing-1999-a}). The final
result is that the unknown moments can be expressed as a linear combination
of the known, with constant coefficients%
\footnote{There are also additional correction terms that appear in the equations
of motion for the moments because of the truncation. These emerge
by deriving the equations of motion from an effective Lagrangian.
This work will be described in a forthcoming paper.%
}. An alternative approach (that has \emph{not} been investigated with
CEID) has been used in a related method\cite{prezhdo-2000-b}.

To use this method with popular approximate electronic structure methods
(such as Hartree-Fock or tight binding - and possibly density functional
theory, though there are additional theoretical problems in this case)
it is necessary to reduce the N-electron problem into a 1-electron
problem (that is, an extension of Hartree-Fock theory). This is achieved
by taking the $N$-electron equations of motion (Eq. \ref{eq:same04})
and taking a trace over all the electrons except one. This produces
equations of motion for 1-electron operators. However, in those equations
of motion 2-electron operators appear. To reduce the theory to a pure
1-electron theory we need to replace these 2-electron operators with
suitable functions of 1-electron operators. This is based on the density
matrix product used in the Hartree-Fock approximation. In matrix notation
the product is $\rho(12,1'2')\approx\rho(1,1')\rho(2,2')-\rho(1,2')\rho(2,1')$,
where the numbers 1 and 2 refer to the set of indices for electrons
1 and 2 respectively. However, this equation by itself is not enough.
There are two reasons: first, we need to approximate 2-electron moments
as well as the electronic density matrix\cite{horsfield-2004-b};
second, since the electron density matrix used in CEID involves a
trace over multiple nuclear configurations, we need to take an appropriate
average of products of single particle density matrices, even for
the electron density matrix\cite{horsfield-2005-a}. In both cases
the starting point is the same, namely to write the moment ($\hat{q}$)
we are interested in terms of the product of nuclear fluctuations
($\hat{Q}$) in the following way\begin{eqnarray}
\hat{q} & = & {\rm Tr}_{N}\left\{ \hat{Q}\hat{\rho}\right\} \nonumber \\
 & = & \int Q(\vec{R}'\vec{R})\hat{\rho}(\vec{R}\vec{R'})\ {\rm d}\vec{R}\ {\rm d}\vec{R}'\nonumber \\
 & = & \int Q(\vec{R}'\vec{R})\rho_{N}(\vec{R}\vec{R'})\hat{\rho}_{e}(\vec{R}\vec{R'})\ {\rm d}\vec{R}\ {\rm d}\vec{R}'\label{eq:SAME06}\end{eqnarray}
 where $\hat{\rho}(\vec{R}\vec{R'})=\langle\vec{R}|\hat{\rho}|\vec{R}'\rangle$,
$\rho_{N}(\vec{R}\vec{R'})={\rm Tr}_{e}\left\{ \hat{\rho}(\vec{R}\vec{R'})\right\} $
and $\hat{\rho}_{e}(\vec{R}\vec{R'})\times\rho_{N}(\vec{R}\vec{R'})=\hat{\rho}(\vec{R}\vec{R'})$.
The operator $\hat{\rho}_{e}(\vec{R}\vec{R'})$ is an electronic density
matrix to which we apply the Hartree-Fock approximation. To evaluate
the integrals over nuclear coordinates we need to make an approximation
for the $\vec{R}$ dependence of $\hat{\rho}_{e}(\vec{R}\vec{R'})$.
We make a Taylor expansion in powers of $\Delta\vec{R}$ about $\hat{\rho}_{e}(\bar{R}\bar{R})$,
which corresponds to weak dynamic coupling between the electrons and
nuclei. For details see references\cite{horsfield-2004-b,horsfield-2005-a}.
Here we briefly consider the the result of the above procedure for
the electronic density matrix:

\begin{eqnarray}
\fl\rho_{e}^{(2)}(12;1'2') & \approx & \rho_{e}^{(1)}(11')\rho_{e}^{(1)}(22')-\rho_{e}^{(1)}(12')\rho_{e}^{(1)}(21')+\nonumber \\
\fl &  & \sum_{\nu\nu'}D_{\nu\nu'}^{RR}\left(\mu_{1,\nu'}^{(1)}(11')\mu_{1,\nu}^{(1)}(22')-\mu_{1,\nu'}^{(1)}(12')\mu_{1,\nu}^{(1)}(21')\right)+\dots\label{eq:xhfeqn}\end{eqnarray}
where $\rho_{e}^{(2)}(12;1'2')$, $\rho_{e}^{(1)}(11')$ and $\mu_{1,\nu'}^{(1)}(11')$
are the matrix representations of the two-electron density matrix,
the one electron density matrix and the one electron first moment
respectively. The matrix $D_{\nu\nu'}^{RR}$ is the inverse of $C_{\nu\nu'}^{RR}={\rm Tr}_{e}\left\{ \hat{\mu}_{2,\nu\nu'}\right\} $.
We see that the density matrix is not quite idempotent even for noninteracting
electrons, which is because there are electron-ion correlations and
a dynamical response of the electrons to nuclear fluctuations. This
response, described by the second term, screens the ion-ion interactions.
The resultant dynamical stiffness corrections are essential for getting
the correct phonon structure and inelastic electron-phonon spectrum\cite{todorov-2005-a,horsfield-2005-a}.

Historically, CEID was developed to allow electric current induced
heating in nanoscale devices to be modelled. Therefore, open boundaries
have been a consideration from the beginning, and have now been implemented
in two ways. The first starts from Eqs. \ref{eq:same00} and \ref{eq:same04}
given above\cite{horsfield-2004-c,bowler-2005-a}. To produce a finite
system with open boundaries we consider our system as being embedded
in an infinite environment, and apply our equations of motion to the
infinite combined system. Clearly we cannot explicitly evolve the
electron density matrix and the moments for an infinite system, so
we make use of the fact that we can write down analytic solutions
for the evolution of the electron density matrix and the moments provided
that the Hamiltonian does not vary with time. This allows us to write
down closed form solutions for the environment which we then couple
to the explicit time evolution of the system in which we are interested.
To produce numerically stable solutions it is found necessary to introduce
a small amount of damping into the environment. For details see references
\cite{horsfield-2004-c,bowler-2005-a}. The second scheme, which is
currently under development, achieves numerical simplifications at
the expense of additional approximations.

\textcolor{black}{So how does this method compare with others? We
have already discussed its relationship to the Ehrenfest approximation,
and clearly it introduces those essential features to the theory that
allow a proper transfer of energy between electrons and nuclei. From
Eq. \ref{eq:same05} we can see some correspondence with surface hopping
(see section \ref{sub:Surface-hopping-methods}), but with the important
difference that different trajectories are able to remain coherent
with one another as they are treated simultaneously rather than independently.
A further important feature is that this method reproduces results
of Fermi's golden rule for the exchange of energy between electrons
are nuclear provided that the second moment is retained\cite{horsfield-2005-a}.}
\textcolor{black}{\emph{}}\textcolor{black}{More broadly, the strengths
of CEID are that it is not perturbative, it does not invoke the notion
of phonons and is inherently anharmonic, it avoids classical interpretations
of quantum transitions, and it in principle contains the mutual screening
of the three interactions (electron-electron, electron-nucleus, nucleus-nucleus),
while retaining the conceptual framework of molecular dynamics.}

CEID is still a rather young method, so a limited range of results
has so far been produced. It has been applied to Joule heating in
atomic wires\cite{horsfield-2004-b}, in which case it was found that
the first moments ($\hat{\mu}_{1,\nu}$ and $\hat{\lambda}_{1,\nu}$)
were essential for producing the correct heating. The second moments
(which involve $\Delta\hat{R}_{\nu}\Delta\hat{R}_{\nu'}$, $\Delta\hat{P}_{\nu}\Delta\hat{R}_{\nu'}$
and $\Delta\hat{P}_{\nu}\Delta\hat{P}_{\nu'}$) are needed to describe
the change in electrical resistance due to increased nuclear vibrations\cite{horsfield-2005-a,todorov-2005-a}.
Once these were included, it was possible to reproduce the spectral
signature of the inelastic scattering of electrons by nuclei.

\section{Conclusion and future directions}

Above we have surveyed some of the phenomena resulting from, and the
current state of theories for understanding, the exchange of energy
between electrons and nuclei. Accurate modelling of the phenomena
is a hard problem because of its intrinsically many-body and correlated
nature. As a consequence, all the theories are in need of further
development, as has been indicated in the text. As the different theories
have different attributes (some are perturbative, others are based
on molecular dynamics, and so on), they can naturally be applied to
different problems. So there is a need for development on more than
one front.

The biggest advance is probably going to be in the range of problems
to be addressed using these techniques. There are now standard phenomena
that are studied or used to test novel methods, such as heating in
nanocontacts, or photodissociation, but they are limited in scope.
Non-equilibrium phenomena, by contrast, are ubiquitous. Certainly
the modelling of molecular electronic devices and biological molecules
associated with non-equilibrium electrons (such as DNA and retinal)
have already begun to be investigated, and their importance must surely
increase. Other problems that should benefit from new methods include
the evolution of radiation damage and inelastic tunneling spectroscopy
carried out with STMs.

\ack C.G.S. is grateful to CONICET for support. DRB is supported
by the Royal Society. This study was partly performed through Special
Coordination Funds for Promoting Science and Technology from the MEXT,
Japan. APH is supported by the IRC in Nanotechnology. HN would like
to thank Th. Martin and M. Brandbyge for illumating and enriching
discussions about non-equilibrium Green's functions. Continued research
on CEID is being supported by the EPSRC, grant numbers EP/C524381/1,
EP/C006739/1 and EP/C524403/1.

\appendix

\section{\label{sec:The-electron-phonon-Hamiltonian}Electron-phonon Hamiltonians}

There are two common approaches to building Hamiltonians that couple
electronic and nuclear dynamics (adiabatic states and static lattice
states). Even though they have been shown to give formally the same
answers in lowest order perturbation theory \cite{burt-1982-a}, there
may be practical reasons for choosing one over the other. We summarise
the basic equations for each case below. Occasionally it is useful
to use the diabatic representation of the electronic states. This
is obtained from the adiabatic representation discussed below by a
unitary transform that makes the nuclear kinetic energy operator diagonal.

\subsection{Adiabatic states}

The Hamiltonian for a system of electrons and nuclei we write as $\hat{H}=\hat{T}_{N}+\hat{H}_{e}(\hat{R})$,
where $\hat{T}_{N}$ is the nuclear kinetic energy, $\hat{H}_{e}(\hat{R})$
is the Hamiltonian for electrons in the field of the nuclei, and $\hat{R}$
is the position operator for all the nuclei. In the Born-Oppenheimer
(adiabatic) separation, we assume that we can neglect the nuclear
kinetic energy to begin with. This then produces the following Schr\"odinger
equation for the electrons in the field of the nuclei\begin{equation}
\hat{H}_{e}(\vec{R})\Phi_{n}(\vec{R})=E_{n}(\vec{R})\Phi_{n}(\vec{R})\label{eq:BO01}\end{equation}
Note that nuclear positions are now represented by numbers $\vec{R}$,
and not operators. The subscript $n$ indexes the Born-Oppenheimer
surfaces. Now that we have a set of electronic states $\Phi_{n}(\vec{R})$,
adiabatic nuclear states $\chi_{nN}$ can be generated by treating
$E_{n}(\vec{R})$ as an effective potential for the nuclei, giving\begin{equation}
\left(\hat{T}_{N}+E_{n}(\vec{R})\right)\chi_{nN}=U_{nN}\chi_{nN}\label{eq:BO02}\end{equation}
where the subscript $N$ indexes the allowed nuclear states, and $U_{nN}$
is the total adiabatic energy (electrons and nuclei). Equipped with
these states, we can now use Fermi's Golden Rule to find transition
rates between the product states $\Psi_{nN}=\chi_{nN}\Phi_{n}$, giving
the matrix elements $M_{nNn'N'}=\int{\rm d}\vec{R}{\rm d}\vec{r}\,\Psi_{nN}^{*}\hat{H}\Psi_{n'N'}$\begin{eqnarray}
\fl M_{nNn'N'} & = & U_{nN}\delta_{nn'}\delta_{NN'}\nonumber \\
\fl & + & \int{\rm d}\vec{R}\,\chi_{nN}^{*}\chi_{n'N'}\left[\int{\rm d}\vec{r}\,\Phi_{n}^{*}\hat{T}_{N}\Phi_{n'}+\sum_{\nu}\frac{\hat{P}_{\nu}\chi_{n'N'}}{M_{\nu}\chi_{n'N'}}\int{\rm d}\vec{r}\,\Phi_{n}^{*}\hat{P}_{\nu}\Phi_{n'}\right]\label{eq:BO03}\end{eqnarray}
where $\vec{r}$ is the set of electronic positions, the subscript
$\nu$ indexes the individual nuclear coordinates, $\hat{P}_{\nu}$
is the corresponding nuclear momentum operator, and $M_{\nu}$ the
nuclear mass. The energies appearing in Fermi's Golden Rule are $U_{nN}$.
If we assume that the energy surfaces on which the nuclei are moving
are harmonic, then the nuclear wave functions are the simple harmonic
oscillator wave functions characterised by a frequency, a mass, and
an equilibrium position. Further, the energies will satisfy\begin{equation}
U_{nN}=E_{n}(\vec{R}_{n,0})+\sum_{\alpha}\left(n_{N\alpha}+\frac{1}{2}\right)\hbar\omega_{n,\alpha}\label{eq:BO04}\end{equation}
where $\vec{R}_{n,0}$ corresponds to the equilibrium positions of
nuclei on surface $n$, and $\alpha$ is an index running over the
normal modes which have angular frequencies $\omega_{n,\alpha}$ and
occupancies $n_{N\alpha}$. The surface dependence of the vibrational
frequencies is often neglected, that is $\omega_{n,\alpha}=\omega_{0,\alpha}$.

\subsection{Static lattice states}

This approach starts with the following partitioning of the Hamiltonian:
\begin{eqnarray}
\fl\hat{H} & = & \hat{T}_{N}+\hat{H}_{e}(\vec{R})\nonumber \\
\fl & = & [\hat{T}_{N}+E_{0}(\vec{R})-E_{0}(\vec{R}_{0})]+\hat{H}_{e}(\vec{R}_{0})+[(\hat{H}_{e}(\vec{R})-E_{0}(\vec{R}))-(\hat{H}_{e}(\vec{R}_{0})-E_{0}(\vec{R}_{0}))]\nonumber \\
\fl & = & \hat{H}_{N}+\hat{H}_{e}(\vec{R}_{0})+\hat{H}_{eN}\label{eq:RL01}\end{eqnarray}
 where $\vec{R}_{0}$ is the equilibrium positions of the nuclei in
the ground state, and $E_{0}(\vec{R})$ is the ground state Born-Oppenheimer
energy surface. The electronic states $\Phi_{n}$ are given by\begin{equation}
\hat{H}_{e}(\vec{R}_{0})\Phi_{n}=E_{n}\Phi_{n}\label{eq:RL02}\end{equation}

There are two important limits in which we can define the states of
the nuclear subsystem. When the electron-nuclear coupling $\hat{H}_{eN}$
is \emph{always} weak then the reference nuclear wavefunctions $\chi_{N}$
are found from\begin{equation}
\hat{H}_{N}\chi_{N}=W_{N}\chi_{N}\label{eq:RL03}\end{equation}
 Fermi's Golden Rule can then be used to find transition rates between
the product states $\Psi_{nN}=\chi_{N}\Phi_{n}$, giving the matrix
elements\begin{equation}
M_{nNn'N'}=\left(E_{n}+W_{N}\right)\delta_{NN'}\delta_{nn'}+\int{\rm d}\vec{R}{\rm d}\vec{r}\,\chi_{N}^{*}\Phi_{n}^{*}\hat{H}_{eN}\chi_{N'}\Phi_{n'}\label{eq:RL04}\end{equation}
In the limit of small ionic displacements we can use a linear approximation
for the interaction Hamiltonian, which is $\hat{H}_{eN}\approx(\vec{R}-\vec{R}_{0})\cdot\vec{\nabla}\hat{H}_{e}(\vec{R}_{0})$,
where we have used the definition of the fixed sites, namely $\vec{\nabla}E_{0}(\vec{R}_{0})=0$.
Matrix elements of this Hamiltonian have the form \begin{equation}
\fl\int{\rm d}\vec{R}{\rm d}\vec{r}\,\chi_{N}^{*}\Phi_{n}^{*}\hat{H}_{eN}\chi_{N'}\Phi_{n'}=\int{\rm d}\vec{R}\,\chi_{N}^{*}(\vec{R}-\vec{R}_{0})\chi_{N'}\cdot\int{\rm d}\vec{r}\,\Phi_{n}^{*}\vec{\nabla}\hat{H}_{e}(\vec{R}_{0})\Phi_{n'}\label{eq:RL05}\end{equation}
The energies appearing in Fermi's Golden Rule are $U_{nN}=E_{n}+W_{N}$.
If the ground state Born-Oppenheimer energy surface is harmonic, then
the nuclear wave functions will be simple harmonic oscillator wave
functions centred about the ground state equilibrium positions. The
nuclear energies will satisfy\begin{equation}
W_{N}=\sum_{\alpha}\left(n_{N\alpha}+\frac{1}{2}\right)\hbar\omega_{\alpha}\label{eq:RL06}\end{equation}
where $\alpha$ is an index running over the normal modes which have
angular frequencies $\omega_{\alpha}$ and occupancies $n_{N\alpha}$.

The second case corresponds to the electron-nuclear coupling being
weak only \emph{between} energy surfaces, but strong on a surface.
In this case the minimum energy configurations of different adiabatic
energy surfaces are displaced significantly from one another. If we
treat the electron-nuclear coupling $\hat{H}_{eN}$ with the linear
approximation we then partition $\hat{F}=-\vec{\nabla}\hat{H}_{e}(\vec{R}_{0})$
into diagonal ($\hat{F}_{D}$) and off-diagonal ($\hat{F}_{OD}$)
terms, where $\int{\rm d}\vec{r}\,\Phi_{n}^{*}\hat{F}_{D}\Phi_{n'}=\delta_{nn'}\int{\rm d}\vec{r}\,\Phi_{n}^{*}\hat{F}\Phi_{n'}$
and $\int{\rm d}\vec{r}\,\Phi_{n}^{*}\hat{F}_{OD}\Phi_{n'}=\left(1-\delta_{nn'}\right)\int{\rm d}\vec{r}\,\Phi_{n}^{*}\hat{F}\Phi_{n'}$.
The diagonal Schr\"odinger equation then becomes$\left(\hat{H}_{N}+\hat{H}_{e}(\vec{R}_{0})-\hat{F}_{D}\cdot(\vec{R}-\vec{R}_{0})\right)\Phi_{n}\tilde{\chi}_{nN}=\tilde{U}_{nN}\Phi_{n}\tilde{\chi}_{nN}$,
where now the nuclear states $\tilde{\chi}_{nN}$ depend on electronic
state. If we make the Harmonic approximation for the nuclear Hamiltonian
$\hat{H}_{N}=\hat{T}_{N}+\frac{1}{2}\vec{u}\cdot{\bf K}\cdot\vec{u}$,
where ${\bf K}$ is the matrix of spring constants and $\vec{u}=\vec{R}-\vec{R}_{0}$
is the displacement of the nuclei relative to the reference positions,
then we get\begin{equation}
\fl\left(\hat{T}_{N}+\frac{1}{2}(\vec{u}-\vec{u}_{n})\cdot{\bf K}\cdot(\vec{u}-\vec{u}_{n})+E_{n}-\frac{1}{2}\vec{u}_{n}\cdot{\bf K}\cdot\vec{u}_{n}\right)\tilde{\chi}_{nN}=\tilde{U}_{nN}\tilde{\chi}_{nN}\label{eq:RL07}\end{equation}
where the displacement $\vec{u}_{n}$ is defined by ${\bf K}\cdot\vec{u}_{n}=\int\Phi_{n}^{*}\hat{F}_{D}\Phi_{n}\,{\rm d}\vec{r}$.
This is just the equation of motion for a shifted oscillator whose
potential minimum is at $\vec{u}_{n}$, and whose energy at the minimum
is $E_{n,0}=E_{n}-\frac{1}{2}\vec{u}_{n}\cdot{\bf K}\cdot\vec{u}_{n}$.
The oscillator wavefunctions are just given by $\tilde{\chi}_{nN}(\vec{u})=\chi_{nN}(\vec{u}-\vec{u}_{n})$.
The energies appearing in Fermi's Golden rule are now\begin{equation}
\tilde{U}_{nN}=E_{n,0}+\sum_{\alpha}\left(n_{N\alpha}+\frac{1}{2}\right)\hbar\omega_{\alpha}\label{eq:RL08}\end{equation}
and the matrix elements are\begin{equation}
\fl M_{nNn'N'}=-(1-\delta_{nn'})\int\Phi_{n}^{*}\hat{F}\Phi_{n'}\,{\rm d}\vec{r}\cdot\int\chi_{N}^{*}(\vec{u}-\vec{u}_{n})\vec{u}\chi_{N'}(\vec{u}-\vec{u}_{n'})\,{\rm d}\vec{R}\label{eq:RL09}\end{equation}

\section{\label{sec:The-Wigner-transform}The Wigner transform}

CEID is formulated in terms of the density matrix, which is then characterised
by moments of the position and momentum fluctuations of the ions.
These moments are reminiscent of moments of classical phase space
distributions, and so it is natural to seek a formal connection. This
can be achieved by appealing to the Wigner matrix\cite{wigner-e-1932,hillery-1984-a,jacoboni-bordone-2004}
which is constructed by applying a transformation to the density matrix
$\hat{\rho}$ which is a function of both electronic and nuclear degrees
of freedom. Let us use a real space representation ($|\vec{X}\rangle$)
of the $N$ nuclear degrees of freedom, and leave the electronic ones
abstract, giving $\hat{\rho}(\vec{X},\vec{X}')=\langle\vec{X}|\hat{\rho}|\vec{X}'\rangle$.
If we make the following linear combinations, $\vec{R}=(\vec{X}+\vec{X}')/2$
and $\vec{S}=\vec{X}-\vec{X}'$, and carrier out a Fourier transformation
with respect to $\vec{S}$, we get the Wigner matrix\begin{equation}
\hat{\rho}_{W}(\vec{R},\vec{P})=\frac{1}{h^{N}}\int\hat{\rho}(\vec{R}+\frac{1}{2}\vec{S},\vec{R}-\frac{1}{2}\vec{S})\exp(\vec{S}\cdot\vec{P}/{\rm i}\hbar)\,{\rm d}\vec{S}\label{eq:WM01}\end{equation}
The connection between this function and a classical phase space distribution
can be seen from the following

\begin{enumerate}
\item If we integrate $\hat{\rho}_{W}(\vec{R},\vec{P})$ over $\vec{P}$,
we get back the quantum spatial distribution function $\hat{\rho}(\vec{R},\vec{R})$.
\item If we integrate $\hat{\rho}_{W}(\vec{R},\vec{P})$ over $\vec{R}$,
we get the quantum momentum distribution function $\hat{\rho}(\vec{P},\vec{P})$.
\item It has an equation of motion similar to that of the classical Liouville
equation in the classical limit defined by $\hbar\to0$, which also
corresponds to heavy nuclei\cite{kapral-1999-a}\begin{eqnarray*}
\fl\frac{\partial\hat{\rho}_{W}(\vec{R},\vec{P})}{\partial t} & = & \frac{1}{{\rm i}\hbar}\left[\hat{H}_{e}(\vec{R}),\hat{\rho}_{W}(\vec{R},\vec{P})\right]\\
\fl & - & \frac{\vec{P}}{M}\cdot\frac{\partial\hat{\rho}_{W}(\vec{R},\vec{P})}{\partial\vec{R}}+\frac{1}{2}\left\{ \frac{\partial\hat{H}_{e}(\vec{R})}{\partial\vec{R}}\cdot\frac{\partial\hat{\rho}_{W}(\vec{R},\vec{P})}{\partial\vec{P}}+\frac{\partial\hat{\rho}_{W}(\vec{R},\vec{P})}{\partial\vec{R}}\cdot\frac{\partial\hat{H}_{e}(\vec{R})}{\partial\vec{R}}\right\} \\
\fl & + & \mathcal{O}(\hbar)\end{eqnarray*}

\end{enumerate}

\section{\label{sec:Equilibrium-and-non-equilibrium}Equilibrium and non-equilibrium
Green's functions}

In this appendix we give the definition of the different Green's functions,
we also briefly introduce the technique of the Keldysh time-contour
applied to non-equilibrium conditions, and derive the corresponding
Green's functions and self-energies in the presence of interactions.

\subsection{\label{sub:General-definitions}General definitions}

Let us consider two operators $A$ and $B$. The retarded $r$, advanced
$a$, time-ordered $t$ and antichronological time-ordered $\tilde{t}$
Green's function with real time arguments are defined as \begin{eqnarray}
\fl G_{A;B}^{r}(t,t') & = & -{\rm i}\theta(t-t')\langle[A(t),B(t')]_{\pm}\rangle\,,\nonumber \\
\fl G_{A;B}^{a}(t,t') & = & {\rm i}\theta(t'-t)\langle[A(t),B(t')]_{\pm}\rangle\,,\nonumber \\
\fl G_{A;B}^{t}(t,t') & = & -{\rm i}\langle T_{t}\left(A(t)B(t')\right)\rangle=-{\rm i}\theta(t-t')\langle A(t)B(t')\rangle\pm{\rm i}\theta(t'-t)\langle B(t')A(t)\rangle\,,\nonumber \\
\fl G_{A;B}^{\tilde{t}}(t,t') & = & -{\rm i}\langle T_{\tilde{t}}\left(A(t)B(t')\right)\rangle=-{\rm i}\theta(t'-t)\langle A(t)B(t')\rangle\pm{\rm i}\theta(t-t')\langle B(t')A(t)\rangle\,.\label{app_defGF}\end{eqnarray}
 The average $\langle\dots\rangle$ is taken over the many-body ground
state and $A(t)$ is given in the Heisenberg representation. The $+$
sign applies when the operators $A$ and $B$ satisfy fermion anticommutation
relations, and the $-$ sign applies if $A$ and $B$ are boson operators.
For some applications it is useful to consider Green's functions without
time ordering. They are the so-called greater $>$ and lesser $<$
Green's functions: \[
\fl G_{A;B}^{>}(t,t')=-{\rm i}\langle A(t)B(t')\rangle\,\,\textrm{and}\,\, G_{A;B}^{<}(t,t')=\pm{\rm i}\langle B(t')A(t)\rangle\,,\]
 with the same sign convention as above. All the other Green's functions
can be defined in terms of these two Green's functions as \begin{eqnarray}
G_{A;B}^{r}(t,t') & = & \theta(t-t')[G_{A;B}^{>}(t,t')-G_{A;B}^{<}(t,t')]\,,\nonumber \\
G_{A;B}^{a}(t,t') & = & \theta(t'-t)[G_{A;B}^{<}(t,t')-G_{A;B}^{>}(t,t')]\,,\nonumber \\
G_{A;B}^{t}(t,t') & = & \theta(t-t')G_{A;B}^{>}(t,t')+\theta(t'-t)G_{A;B}^{<}(t,t')\,.\label{app_defGFbis}\end{eqnarray}
 At (thermal) equilibrium or in a stationary state regime, the Green's
functions depend on the time difference only (\emph{i.e.} $G_{A;B}^{x}(t,t')=G_{A;B}^{x}(t-t')$),
and their Fourier transform is dependent on only one energy argument
$G^{x}(\omega)$. The advanced and retarded Green's functions $G^{a,r}(\omega)$
contain information about the spectral density of the system, while
the lesser and greater Green's functions $G^{<,>}(\omega)$ contain
information about both the spectral density and the occupancy of the
system at or out of equilibrium.

\subsection{\label{sub:Non-equilibrium-conditions-and}Non-equilibrium conditions
and time loop contour}

Here we briefly explain the principles of non-equilibrium Green's
function \cite{LKeldysh65,LKadanoff1962}. Consider a many particle
system with interactions and/or with a coupling to an external driving
force (external field). We want to study the system by reducing its
mathematical description to the calculation of a perturbation series,
with the hope that later on we can resum some (if not all) the contributions,
as is usually done in many-body statistical physics \cite{GMahan1990,AAbrikosov1963,AFetter1971}.
Therefore, we start from the non-interacting ground state at the infinitely
remote past (where there are no interactions) and the interaction
$V$ is turned on adiabatically. Then the different Green's functions
are calculated by going to the interaction picture and evaluating
the terms of the perturbation expansion series with Wick's theorem.
This theorem applies to a time-ordered average with respect to a Hamiltonian
that is quadratic in creation/annihilation operators (for example,
a non-interacting Hamiltonian). According to Wick's theorem, the average
of any product of operators can be found by forming all pairs of operators
and replacing these by their average.

For a non-equilibrium (interacting) system, the ground state in the
future is not known \emph{a} \emph{priori}, and we are left with average
products which are only partially time ordered. The Keldysh recipe
is to introduce a time contour along which the operators can be ordered.
The time contour $C_{K}$ contains two branches, the upper $(+)$
and the lower $(-)$ branch. On the upper branch, time starts in the
infinitely remote past and evolves forwards, then at the turning point
(which can be placed at any arbitrary time), one passes onto the lower
branch where the system evolves backwards in time back to the initially
non-interacting starting point at $t=-\infty$. Then any expectation
value of products of operator reduces to $\langle\phi_{0}|T_{C_{K}}\left(\hat{A}(t)\hat{B}(t')\dots S_{C_{K}}\right)|\phi_{0}\rangle$
where $\langle\phi_{0}|\dots|\phi_{0}\rangle$ is the average over
the non-interacting ground state. The operators are given in the interaction
picture, \emph{i.e.} $\hat{A}(t)={\rm e}^{{\rm i}H_{0}t/\hbar}A\,\,{\rm e}^{-{\rm i}H_{0}t/\hbar}$.
$S_{C_{K}}$ is the generalization of the time evolution operator
(Eq. \ref{Uttp}) on the time loop contour $S_{C_{K}}=T_{C_{K}}\left(\exp\{-{\rm i}/\hbar\int_{C_{K}}{\rm d}\tau\hat{V}(\tau)\}\right)$
where $T_{C_{K}}$ is the time-ordering operator on the contour $C_{K}$
and $\int_{C_{K}}{\rm d}t$ implies integration over $C_{K}$. With
these definitions, any time ordered product can be calculated using
the usual rules of many-body perturbation theory (Feynman diagrammatic
expansion, Wick's theorem, \emph{etc}.) \cite{GMahan1990,AAbrikosov1963,AFetter1971}.
The Keldysh recipe is equivalent to reducing the problem to the calculation
of averages over the non-interacting ground state, which is a great
achievement because such averages can be calculated exactly in a lot
of cases \cite{JRammer86}. However, there is a price to pay for that:
now we have to work with four different Green's functions defined
by the position of the two times $(t,t')$ on $C_{K}$. When the two
times $(t,t')$ are on the same branch, the time ordering $T_{C_{K}}$
is equivalent to the standard time ordering: forward time ordering
on the upper branch and backward time (or anti-time) ordering on the
lower branch. When $(t,t')$ are on different branches, the time ordering
is such that any time on the lower branch is always later on the time
loop contour $C_{K}$ than any time on the upper branch.

The electron Green's function defined from the fermion operator $\Psi$
(with the definitions in section \ref{sub:General-definitions}: $A=\Psi$
and $B=\Psi^{\dagger}$) $G(t,t')=-{\rm i}\langle T_{C_{K}}\left(\Psi(t)\Psi^{\dagger}(t')\right)\rangle$
has four Keldysh components on $C_{K}$: for $(t,t')$ on the upper
branch $(+)$ $G(t,t')=-{\rm i}\langle T_{C_{K}}\left(\Psi(t_{+})\Psi^{\dagger}(t'_{+})\right)\rangle=G^{t}(t,t')$;
for $(t,t')$ on the lower branch $(-)$ $G(t,t')=-{\rm i}\langle T_{C_{K}}\left(\Psi(t_{-})\Psi^{\dagger}(t'_{-})\right)\rangle=G^{\tilde{t}}(t,t')$;
for $t$ on the $(+)$ branch and $t'$ on the $(-)$ branch $G(t,t')=-{\rm i}\langle T_{C_{K}}\left(\Psi(t_{+})\Psi^{\dagger}(t'_{-})\right)\rangle=G^{<}(t,t')$;
and finally for $t$ on the $(-)$ branch and $t'$ on the $(+)$
branch $G(t,t')=-{\rm i}\langle T_{C_{K}}\left(\Psi(t_{-})\Psi^{\dagger}(t'_{+})\right)\rangle=G^{>}(t,t')$.
However these Green's functions are not completely independent. They
satisfy the following relations: $G^{t}+G^{\tilde{t}}=G^{<}+G^{>}$
and $G^{r}-G^{a}=G^{>}-G^{<}$ (with the retarded and advanced Green's
functions defined as in section \ref{sub:General-definitions} with
$A=\Psi$ and $B=\Psi^{\dagger}$). The self-energy $\Sigma$, associated
with the interaction $V$ for example, also has four components on
the contour $C_{K}$.

It can be shown that the Green's function in the presence of interaction
$G$ is related to the Green's function in the absence of interaction
$G_{0}$ via the usual Dyson equation $G(t,t')=G_{0}(t,t')+\int_{C_{K}}{\rm d}t_{1}{\rm d}t_{2}\, G_{0}(t,t_{1})\Sigma(t_{1},t_{2})G(t_{2},t')$
where the time integrals are taken over $C_{K}$. The Dyson equation
can then be re-expressed using only integrals over the real-time axis
by introducing the different components of $G$ and $\Sigma$ on the
contour $C_{K}$. By using the relation between the different Green's
functions, we finally obtain a Dyson-like equation for the advanced
and retarded (non-equilibrium) Green's functions: $G^{r,a}=G_{0}^{r,a}+G_{0}^{r,a}\Sigma^{r,a}G^{r,a}$
and another kinetic equation for the (non-equilibrium) lesser and
greater Green's functions: $G^{<,>}=(1+G^{r}\Sigma^{r})G_{0}^{<,>}(1+\Sigma^{a}G^{a})+G^{r}\Sigma^{<,>}G^{a}$.
In these equations, the products $G\Sigma$ or $\Sigma G$ imply time
integration over the real axis, \emph{i.e.} $\int_{-\infty}^{+\infty}{\rm d}t$.
Similar results can be derived for the phonon Green's functions (\emph{i.e.}
when $A$ and $B$ are phonon field operators).

In principle, we can now calculate exactly the non-equilibrium properties
of a many-body interacting system by determining self-consistently
the different Green's functions and self-energies (knowing that the
self-energies are functionals of the different Green's functions themselves).

\subsection{\label{sub:Electron-and-phonon}Electron and phonon Green's functions}

\subsubsection{\label{sub:Non-interacting-systems}Non-interacting systems}

For quadratic Hamiltonians (\emph{i.e.} non-interacting particles),
we can calculate exactly the different Green's functions. Let us start
with electrons and consider the electronic Hamiltonian $H_{0}=\sum_{n}\epsilon_{n}c_{n}^{\dagger}c_{n}$.
The operator $c_{n}^{\dagger}$ ($c_{n}$) creates (annihilates) an
electron in the $n$-th electronic state with energy $\epsilon_{n}$;
$n$ labels either the eigenstates of a finite size system or the
$k$-states of a (1,2,3)-dimensional periodic system. With the definitions
given in the previous sections and taking $A=c_{n}$ and $B=A^{\dagger}$,
we find the following Green's functions in the energy representation:
\begin{eqnarray}
\fl G_{0}^{r,a}(\omega) & = & \frac{1}{\omega-\epsilon_{n}\pm{\rm i}\eta}\,\,\textrm{with }\eta\rightarrow0^{+}\,,\nonumber \\
\fl G_{0}^{<}(\omega) & = & 2\pi{\rm i}\langle c_{n}^{\dagger}c_{n}\rangle\,\delta(\omega-\epsilon_{n})\,,\,\,\,\, G_{0}^{>}(\omega)=-2\pi{\rm i}(1-\langle c_{n}^{\dagger}c_{n}\rangle)\,\,\delta(\omega-\epsilon_{n})\,.\label{app_defGw}\end{eqnarray}
 The average $\langle c_{n}^{\dagger}c_{n}\rangle$ gives the occupation
number of the state $n$, and in the thermodynamic limit it is given
by the Fermi-Dirac distribution $f(\epsilon_{n})=1/({\rm e}^{\beta(\epsilon_{n}-\mu)}+1)$
where $\mu$ is chemical potential and $\beta=1/k_{B}T$.

For phonons with a quadratic Hamiltonian $H_{0}=\sum_{\lambda}\hbar\omega_{\lambda}(a_{\lambda}^{\dagger}a_{\lambda}+1/2)$
where $a_{\lambda}^{\dagger}$ ($a_{\lambda}$) creates (annihilates)
a quantum of energy $\hbar\omega_{\lambda}$, we find the following
Green's functions (denoted by the letter $D$) defined from $A=a_{\lambda}^{\dagger}+a_{\lambda}$
and $B=A^{\dagger}$: \begin{eqnarray}
\fl D_{0}^{r,a}(\omega) & = & \frac{1}{\omega-\omega_{\lambda}\pm{\rm i}\eta}-\frac{1}{\omega+\omega_{\lambda}\pm{\rm i}\eta}\,\,\textrm{with }\eta\rightarrow0^{+}\,,\nonumber \\
\fl D_{0}^{<,>}(\omega) & = & -2\pi{\rm i}\{\langle n_{\lambda}\rangle\delta(\omega\mp\omega_{\lambda})+(1+\langle n_{\lambda}\rangle)\,\,\delta(\omega\pm\omega_{\lambda})\}\,,\label{app_defDw}\end{eqnarray}
 where $n_{\lambda}=a_{\lambda}^{\dagger}a_{\lambda}$. In the thermodynamic
limit, $\langle n_{\lambda}\rangle$ is given by the Bose-Einstein
distribution $N(\omega)=1/({\rm e}^{\beta\omega}-1)$.

\subsubsection{\label{sub:Coupling-to-reservoirs}Coupling to reservoirs}

If the (finite size) system of interest is connected to $M$ other
subsystems (generally much larger) which act as either thermal reservoirs
or particle reservoirs, then we can calculate the full Green's function
of the connected primary system (or central region) from the Green's
function of the isolated parts of the system. This is done by solving
the Dyson equation and the quantum kinetic equation for the non-equilibrium
Green's functions. As an example, we consider that the coupling to
the $M$ subsystems is given by the Hamiltonian matrices $V_{\alpha}$
corresponding to electron hopping between the primary system and the
$M$ other subsystems. When neglecting the interactions between particles,
the self-energies $\Sigma^{x=(r,a,<,>)}$ entering the Dyson equations
for the electron Green's functions $G^{x}$ of the central region
(coupled to the $M$ reservoirs) are $\Sigma^{x}=\sum_{\alpha=1}^{M}\Sigma_{\alpha}^{x}$
where $\Sigma_{\alpha}^{x}(\omega)=V_{\alpha}g_{\alpha\alpha}^{x}(\omega)V_{\alpha}$
and $g_{\alpha\alpha}^{x}$ is the corresponding Green's function
of the isolated $\alpha$-th subsystem (Eq.~\ref{app_defGw}).

For a central region consisting only of phonon modes $\lambda$ coupled
to another phonon bath (set of phonon of frequency $\omega_{\beta}$)
via some coupling contants $U_{\beta}$, one can derive the full phonon
Green's function of the central region as $D^{r,a}=[{D_{0}^{r,a}}^{-1}-\Pi_{\lambda}^{r,a}]^{-1}$
where $D_{0}^{r,a}$ is the bare phonon Green's function given in
Eq.~\ref{app_defDw} and the self-energy $\Pi_{\lambda}^{r,a}$ arises
from the coupling of the modes $\lambda$ to the modes $\beta$. In
the simplest case, $\Pi_{\lambda}^{r,a}$ can be approximated by $\Pi_{\lambda}^{r,a}(\omega)\propto{\rm i}\sum_{\beta}|U_{\beta}|^{2}\delta(\omega-\omega_{\beta})$
\cite{MGalperin05}.

In more realistic systems, the central region needs to be described
by a Hamiltonian that also includes interaction between electrons
or between electrons and phonons. In the next section, the case of
electron-phonon interaction is considered in detail.

\subsubsection{\label{sub:Self-energies-for-electron-phonon}Self-energies for electron-phonon
interaction}

A general form for the linear electron-phonon ($e$-ph) interaction
Hamiltonian is as follows: $H_{\textrm{\textit{e}ph}}=\sum_{\lambda,n,m}\gamma_{\lambda nm}(a_{\lambda}^{\dagger}+a_{\lambda})c_{n}^{\dagger}c_{m}$
where $\gamma_{\lambda nm}$ are the matrix elements of the $e$-ph
coupling matrix $\overline{\gamma}_{\lambda}$. For such an interaction,
one needs to include another contribution $\Sigma_{eph}^{x}$ in the
electron self-energies $\Sigma^{x}=\sum_{\alpha=1}^{M}\Sigma_{\alpha}^{x}$.
Similarily, one has to include the contribution due to the $e$-ph
interaction in the phonon self-energies $\Pi_{\lambda}^{x}$.

As mentioned in section \ref{sub:Non-equilibrium-Green's-functions},
within the self-consistent Born approximation, the lowest-order perturbation
expansion for the interaction is used to determine the self-energies
and then the non-interacting Green's functions is substituted by the
full Green's functions of the system. As an example, the contribution
from the Fock-diagrams to the electron self-energy is: $\Sigma^{\textrm{F}}(t,t')\propto{\rm i}\sum_{\lambda}D(t,t')\overline{\gamma}_{\lambda}G(t,t')\overline{\gamma}_{\lambda}$
($t$ and $t'$ being on the contour $C_{K}$). Using Langreth's rules
for products of operators on $C_{K}$ \cite{HHaug1996}, one gets
the different contributions to the self-energy in the following form:
$\Sigma_{e\textrm{ph}}^{xy}(\omega)\propto i\int{\rm d}\omega'D^{x}(\omega-\omega')\overline{\gamma}_{\lambda}G^{y}(\omega')\overline{\gamma}_{\lambda}$,
where $x,y=(r,a,<,>)$. For example, the retarded Fock part of the
electron self-energy is given by: $\Sigma_{e\textrm{ph}}^{\textrm{F},r}=\Sigma_{e\textrm{ph}}^{r<}+\Sigma_{e\textrm{ph}}^{rr}+\Sigma_{e\textrm{ph}}^{<r}$.
The exact expressions for the different self-energies can be found
in Refs. \cite{TFrederiksen04,MPaulsson05,JViljas05,LVega05,AJauho05}.
The contribution to the phonon self-energies $\Pi_{\lambda}^{x}$
can be calculated to second order in the $e$-ph coupling by considering
diagrams for electron-hole excitations. The polarisation $\Pi_{\lambda}$
can be obtained from $\Pi_{\lambda}(t,t')\propto{\rm i}\textrm{Tr}\{\overline{\gamma}_{\lambda}G(t,t')\overline{\gamma}_{\lambda}G(t',t)\}$
where the trace runs over the electron states $n$. Once again, one
has to use Langreth's rules to obtain the different contributions
$\Pi_{\lambda}^{a,r,<,>}(\omega)$ which can be found in Refs.~\cite{TMii02,TMii03,AMitra04,JViljas05}.
And finally, we solve the problem by calculating the different electron
and phonon full Green's functions $G^{a,r,<,>}(\omega)$ and $D_{\lambda}^{a,r,<,>}(\omega)$
from the coupled Dyson and quantum kinetic equations, with the self-energies
being functionals in the following form: $\Sigma_{e\textrm{ph}}^{a,r,<,>}(\omega)=\Sigma[\{ G^{x}(\omega)\},\{ D^{x}(\omega)\}]$
and $\Pi_{\lambda}^{a,r,<,>}(\omega)=\Pi[\{ G^{x}(\omega)\},\{ D^{x}(\omega)\}]$.

\section*{References}

\providecommand{\newblock}{}


\end{document}